\documentclass{llncs}

\usepackage{chngpage}

\usepackage{booktabs}

\usepackage[utf8]{inputenc}

\usepackage{scrextend}

\usepackage[usenames, dvipsnames]{color}

\usepackage{tikz}
\usepackage{url}
\usepackage{amsmath, amssymb}
\usepackage{mathabx}
\usepackage{caption} 
\usepackage{fancyvrb}
\DefineVerbatimEnvironment{ex}{Verbatim}{numbers=left,numbersep=2mm,frame=single,fontsize=\scriptsize}

\usepackage{xspace}

\usepackage{tikz}
\def\checkmark{\tikz\fill[scale=0.4](0,.35) -- (.25,0) -- (1,.7) -- (.25,.15) -- cycle;} 

\usepackage{pifont}

\usepackage{tabularx}

\usepackage[colorinlistoftodos]{todonotes}

\newenvironment{table-1cols}{
  \scriptsize
  \sffamily
  \vspace{0.3cm}
  \begin{tabular}{l}
  \hline
  \textbf{Requirements} \\
  \hline

}{
  \hline
  \end{tabular}
  \linebreak
}

\newenvironment{table-2cols}{
  \scriptsize
  \sffamily
  \vspace{0.3cm}
  \begin{tabular}{l|l}
  \hline
  \textbf{Requirements} & \textbf{Covering DSCLs} \\
  \hline

}{
  \hline
  \end{tabular}
  \linebreak
}

\newenvironment{constraint-languages-complexity}{
  \begin{tabular}{l|c|c|c|c|c|c}
  \hline
  \textbf{Complexity Class} & \textbf{DSP} & \textbf{OWL2-DL} & \textbf{OWL2-QL} & \textbf{ReSh} & \textbf{ShEx} & \textbf{SPIN} \\
  \hline

}{
  \hline
  \end{tabular}
  \linebreak
}

\newenvironment{user-fiendliness}{
  \begin{tabular}{l|c|c|c|c|c}
  \hline
  \textbf{criterion} & \textbf{DSP} & \textbf{OWL2} & \textbf{ReSh} & \textbf{ShEx} & \textbf{SPIN} \\
  \hline

}{
  \hline
  \end{tabular}
  \linebreak
}

\usepackage{array,graphicx}
\usepackage{booktabs}
\usepackage{pifont}
\newcommand*\rot{\rotatebox{90}}

\usepackage{booktabs}

\usepackage{tablefootnote}

\usepackage{float}

\usepackage{multirow}

\begin{document}
\renewcommand{\arraystretch}{1.3}
%
%

\title{Evaluating the Quality of RDF Data Sets on}
\subtitle{Common Vocabularies in the Social, Behavioral, and Economic Sciences}


\titlerunning{XXXXX}  
%
\author{Thomas Hartmann\inst{1} \and Benjamin Zapilko\inst{1} \and Joachim Wackerow\inst{1} \and Kai Eckert\inst{2}}
\authorrunning{} 
%
\institute{GESIS – Leibniz Institute for the Social Sciences, Germany\\
\email{\{firstname.lastname\}@gesis.org},\\ 
\and
University of Mannheim, Germany \\
\email{kai@informatik.uni-mannheim.de} 
}

\maketitle              

\begin{abstract}
From 2012 to 2015 together with other Linked Data community members and experts from the social, behavioral, and economic sciences (\emph{SBE}), we developed diverse vocabularies to represent SBE metadata and tabular data in RDF.
The \emph{DDI-RDF Discovery Vocabulary (DDI-RDF)} 
is designed to support the dissemination, management,
and reuse of unit-record data, i.e., data about individuals, households, and businesses, collected in form of responses to studies and archived for research purposes.
The \emph{RDF Data Cube Vocabulary (QB)} is a W3C recommendation for expressing \emph{data cubes}, i.e. multi-dimensional aggregate data and its metadata. 
\emph{Physical Data Description (PHDD)} is a vocabulary to model data in rectangular format, i.e., tabular data. 
The data could either be represented in records with character-separated values (CSV) or fixed length. 
The \emph{Simple Knowledge Organization System (SKOS)} is a vocabulary to build knowledge organization systems such as thesauri, classification schemes, and taxonomies.
XKOS is a SKOS extension to describe formal statistical classifications. 

To ensure high quality of and trust in both metadata and data, their representation in RDF must satisfy certain criteria - specified in terms of RDF constraints.
In this paper, we evaluate the data quality of 15,694 data sets (4.26 billion triples) of research data for the social, behavioral, and economic sciences obtained from 33 SPARQL endpoints. We checked 115 constraints on three different and representative SBE vocabularies (DDI-RDF, QB, and SKOS) by means of the \emph{RDF Validator}, a validation environment which is available at \url{http://purl.org/net/rdfval-demo}.

\end{abstract}

\section{Introduction}

For constraint formulation and RDF data validation, several languages exist or are currently developed. \emph{Shape Expressions (ShEx)}, \emph{Resource Shapes (ReSh)}, \emph{Description Set Profiles (DSP)}, \emph{OWL 2}, the \emph{SPARQL Inferencing Notation (SPIN)}, and \emph{SPARQL} are the six most promising and widely used constraint languages. OWL 2 is used as a constraint language under the closed-world and unique name assumptions. The W3C currently develops \emph{SHACL}, an RDF vocabulary for describing RDF graph structures. With its direct support of validation via SPARQL, SPIN is very popular and certainly plays an important role for future developments in this field. It is particularly interesting as a means to validate arbitrary constraint languages by mapping them to SPARQL \cite{BoschEckert2014-2}. Yet, there is no clear favorite and none of the languages is able to meet all requirements raised by data practitioners. Further research and development therefore is needed.

In 2013, the W3C organized the RDF Validation Workshop,\footnote{\url{http://www.w3.org/2012/12/rdf-val/}} 
where experts from industry, government, and academia discussed first use cases for constraint formulation and RDF data validation.
In 2014, two working groups on RDF validation have been established to develop a language to express constraints on RDF data: 
the \emph{W3C RDF Data Shapes Working Group}\footnote{\url{http://www.w3.org/2014/rds/charter}} (33 participants of 19 organizations) and the \emph{DCMI RDF Application Profiles Task Group}\footnote{\url{http://wiki.dublincore.org/index.php/RDF-Application-Profiles}} (29 people of 22 organizations) which among others bundles the requirements of data institutions of the cultural heritage sector and the \emph{social, behavioral, and economic (SBE)} sciences and represents them in the W3C group. 

Within the DCMI task group, a collaboratively curated database of RDF validation requirements\footnote{Online available at: \url{http://purl.org/net/rdf-validation}} has been created which contains the findings of the working groups based on various case studies provided by data institutions \cite{BoschEckert2014}. It is publicly available and open for further contributions.
The database connects requirements to use cases, case studies, and implementations and forms the basis of this paper. 
We distinguish 81 requirements to formulate constraints on RDF data; 
each of them corresponding to a constraint type.

We collected constraints for commonly used vocabularies in the SBE domain, either from the vocabularies themselves or from domain and data experts, in order to gain a better understanding about the role of certain requirements for data quality and to direct the further development of constraint languages. All in all, this lead to 115 constraints we implemented on three vocabularies.
We let the experts classify the constraints according to the severity of their violation. 

As we do not want to base our conclusions on the evaluation of vocabularies and constraint definitions alone, we conducted a large-scale experiment.
For all these implemented 115 constraints, we evaluated the data quality of 15,694 data sets (4.26 billion triples) of SBE research data on three common vocabularies in SBE sciences (DDI-RDF, QB, SKOS) obtained from 33 SPARQL endpoints.

\section{Common Vocabularies in SBE Sciences}
\label{sbe-vocabularies}

We took all well-established and newly developed SBE vocabularies into account and defined constraints for three vocabularies commonly used in the SBE sciences which are briefly introduced in the following. We analyzed actual data according to constraint violations, as for these vocabularies large data sets are already published.

SBE sciences require high-quality data for their empirical research. For more than a decade, members of the SBE community have been developing and using a
metadata standard, composed of almost twelve hundred metadata fields, known as the \emph{Data Documentation Initiative (DDI)}, \footnote{\url{http://www.ddialliance.org/Specification/}}
an XML format to disseminate, manage,
and reuse data collected and archived for research \cite{Vardigan-2008}. 
In XML, the definition of schemas containing constraints and the validation of data according to these constraints is commonly used to ensure a certain level of data quality.
With the rise of the Web of Data, data professionals and institutions are very interested in having their data be discovered and used by publishing their data directly in RDF or at least publish accurate metadata about their data to facilitate data integration. Therefore, not only established vocabularies like SKOS are used; 
recently, members of the SBE and Linked Data community developed with the \emph{DDI-RDF Discovery Vocabulary (DDI-RDF)}\footnote{\url{http://rdf-vocabulary.ddialliance.org/discovery.html}} a means to expose \emph{DDI} metadata as Linked Data. 

The data most often used in research within SBE sciences is \emph{unit-record data}, i.e., data collected about individuals, businesses, and households, in form of responses to studies or taken from administrative registers such as hospital records, registers of births and deaths. A \emph{study} represents the process by which a data set was generated or collected. The range of unit-record data is very broad - including census, education, health data and business, social, and labor force surveys. This type of research data is held within data archives or data libraries after it has been collected, so that it may be reused by future researchers. By its nature, unit-record data is highly confidential and access is often only permitted for qualified researchers who must apply for access. Researchers typically represent their results as aggregated data in form of multi-dimensional tables with only a few columns: so-called \emph{variables} such as \emph{sex} or \emph{age}. Aggregated data, which answers particular research questions, is derived from unit-record data by statistics on groups or aggregates such as frequencies and arithmetic means. The purpose of publicly available aggregated data is to get a first overview and to gain an interest in further analyses on the underlying unit-record data. For more detailed analyses, researchers refer to unit-record data including additional variables needed to answer subsequent research questions. 

\emph{Formal childcare} is an example of an aggregated variable which captures the measured availability of childcare services in percent over the population in European Union member states by 
the dimensions \emph{year}, \emph{duration}, \emph{age} of the child, and \emph{country}.
Variables are constructed out of values (of one or multiple datatypes) and/or code lists.
The variable \emph{age}, e.g., may be represented by values of the datatype \emph{xsd:nonNegativeInteger} or by a code list of age clusters (e.g., '0 to 10' and '11 to 20'). 
The \emph{RDF QB Vocabulary (QB)}\footnote{http://www.w3.org/TR/vocab-data-cube/} is a W3C recommendation for representing \emph{QBs}, i.e., multi-dimensional aggregated data, in RDF \cite{Cyganiak2010}. 
A \emph{qb:DataStructureDefinition} contains metadata of the data collection.
The variable \emph{formal childcare} is modeled as \emph{qb:measure}, since it stands for what has been measured in the data collection.
\emph{Year}, \emph{duration}, \emph{age}, and \emph{country} are \emph{qb:dimensions}.
Data values, i.e., the availability of childcare services in percent over the population, are collected in a \emph{qb:DataSet}. 
Each data value is represented inside a \emph{qb:Observation} which contains values for each dimension. 

For more detailed analyses we refer to the underlying unit-record data. The aggregated variable \emph{formal childcare} is calculated on the basis of six unit-record variables (i.a., \emph{Education at pre-school}) for which detailed metadata is given (i.a., code lists) enabling researchers to replicate the results shown in aggregated data tables.
\emph{DDI-RDF} is used to represent metadata on unit-record data in RDF.
The study (\emph{disco:Study}) for which the unit-record data has been collected 
contains eight data sets (\emph{disco:LogicalDataSet})
including variables (\emph{disco:Variable}) like the six ones needed to calculate the variable \emph{formal childcare}.

The \emph{Simple Knowledge Organization System (SKOS)} is reused to a large extend to build SBE vocabularies.
The codes of the variable \emph{Education at pre-school} are modeled as \emph{skos:Concepts} and 
a \emph{skos:OrderedCollection} organizes them in a particular order within a \emph{skos:memberList}.
A variable may be associated with a theoretical concept (\emph{skos:Concept}) and \emph{skos:narrower} builds the hierarchy of theoretical concepts within a \emph{skos:ConceptScheme} of a study.
The variable \emph{Education at pre-school} is assigned to the theoretical concept \emph{Child Care} which is a narrower concept of the top concept \emph{Education}.
Controlled vocabularies (\emph{skos:ConceptScheme}), serving as extension and reuse mechanism,
organize types (\emph{skos:Concept}) of descriptive statistics (\emph{disco:SummaryStatistics}) like minimum, maximum, and arithmetic mean.

\section{Classification of Constraint Types and Constraints}
\label{classification}

To gain better insights into the role that certain types of constraints play for the quality of RDF data, we use two simple classifications: on the one hand, we classify RDF constraint types whether they are expressible by different types of constraint languages and on the other hand, we classify constraints formulated for a given vocabulary according to the perceived severity of their violation. 

Within the working groups, we identified by today 81 requirements to formulate RDF constraints (e.g., \emph{R-75: minimum qualified cardinality restrictions}); each of them corresponding to an RDF constraint type.\footnote{Constraint types and constraints are uniquely identified by alphanumeric technical identifiers like \emph{R-71-CONDITIONAL-PROPERTIES}} 
Within a technical report, we explain each requirement/constraint type in detail and give examples for each expressed by different constraint languages \cite{BoschNolleAcarEckert2015}. We provide mappings to representations in Description Logics (DL) \cite{Baader-2003} to logically underpin each requirement and to determine which DL constructs are needed to express each constraint type.
For the three vocabularies, several SBE domain experts determined the default severity level of the 115 concrete constraints, which we published in a technical report \cite{BoschZapilkoWackerowEckert2015}.
In the following, we summarize the classifications of constraint types and constraints for the purpose of our evaluation.

\subsection{Classification of Constraint Types according to the Expressivity of Constraint Languages}

According to the expressivity of constraint languages, the complete set of constraint types encompasses three not disjoint sets of constraint types:
\begin{enumerate}
	\item \emph{RDFS/OWL Based}
	\item \emph{Constraint Language Based}
	\item \emph{SPARQL Based}
\end{enumerate}

\textbf{\emph{RDFS/OWL Based}.}
The modeling languages RDFS and OWL are typically used to formally specify vocabularies. \emph{RDFS/OWL Based} denotes the set of constraint types which can be formulated with RDFS/OWL axioms which we use in terms of constraints with CWA/UNA semantics and without reasoning.\footnote{The entailment regime is to be decided by the implementers. It is our point that reasoning affects validation and that a proper definition of the reasoning to be applied is needed.} 

\textbf{\emph{Constraint Language Based}.}
We further distinguish \emph{Constraint Language Based} as the set of constraint types that can be expressed by common classical declarative high-level constraint languages like ShEx, ReSh, and DSP. 
There is a strong overlap between \emph{RDFS/OWL} and \emph{Constraint Language Based} constraint types as in many cases constraint types are expressible by both classical constraint languages and OWL. SPARQL, however, is considered as a low-level implementation language in this context. In contrast to SPARQL, high-level constraint languages are comparatively easy to understand and constraints can be formulated more concisely. Declarative languages may be placed on top of SPARQL when using it as an implementation language. 

\textbf{\emph{SPARQL Based}.}
The set \emph{SPARQL Based} encompasses constraint types that are not expressible by RDFS/OWL or common high-level constraint languages but by plain SPARQL. 

\subsection{Classification of RDF Constraints according to the Severity of Constraint Violations}

A concrete constraint is instantiated from one of the 81 constraint types and is defined for a specific vocabulary.
It does not make sense to determine the severity of constraint violations of an entire constraint type,
as the severity depends on the individual context and vocabulary.
SBE experts determined the default \emph{severity level}\footnote{The possibility to define severity levels in vocabularies is in itself a requirement (\emph{R-158}).} for each constraint to indicate how serious the violation of the constraint is. We use the classification system of log messages in software development like \emph{Apache Log4j 2} \cite{Apache-2015}, the \emph{Java Logging API},\footnote{\url{http://docs.oracle.com/javase/7/docs/api/java/util/logging/Level.html}} and the \emph{Apache Commons Logging API}\footnote{\url{http://commons.apache.org/proper/commons-logging/}} as many data practitioners also have experience in software development and software developers intuitively understand these levels. We simplify this commonly accepted classification system and distinguish the three severity levels (1) \emph{informational}, (2) \emph{warning}, and (3) \emph{error}.
Violations of \emph{informational} constraints point to desirable but not necessary data improvements to achieve RDF representations which are ideal in terms of syntax and semantics of used vocabularies. 
\emph{Warnings} are syntactic or semantic problems which typically should not lead to an abortion of data processing.
\emph{Errors}, in contrast, are syntactic or semantic errors which should cause the abortion of data processing. 
Although we provide default severity levels for each constraint, validation environments should enable users to adapt the severity levels of constraints according to their individual needs.

\section{Evaluation}

In this section, we describe our results based on an automatic constraint checking of a large data set. Despite the large volume of the data set in general, we have to keep in mind that this study only uses data for three vocabularies. As described in Section~\ref{sbe-vocabularies}, for other vocabularies there is often not (yet) enough data openly available to draw general conclusions. The three vocabularies, however, are representative, cover different aspects of SBE data, and are also a mixture of widely adopted and accepted well-established vocabularies (QB, SKOS) and a vocabulary under development (DDI-RDF\footnote{Expected publication at the end of the year 2015}).

\subsection{Experimental Setup}
\label{implementation}


On the three vocabularies (DDI-RDF, QB, SKOS), we identified and classified 115 constraints\footnote{All 115 implemented constraints are online available at: \url{https://github.com/boschthomas/rdf-validation/tree/master/constraints}} which we implemented for data validation. We ensured that the implementation of the constraints is equally distributed over the classes and vocabularies we have. We then evaluated the data quality of 15,694 data sets (4.26 billion triples) of SBE research data using these 115 constraints, obtained from 33 SPARQL endpoints.



Table~\ref{tab:datasets} lists the number of validated data sets and the overall sizes in terms of triples for each of the vocabularies. We validated, i.a., 
(1) QB data sets published by the \emph{Australian Bureau of Statistics},
the \emph{European Central Bank}, and the
\emph{Organisation for Economic Co-operation and Development},
(2) SKOS thesauri like the \emph{AGROVOC Multilingual agricultural thesaurus},
the \emph{STW Thesaurus for Economics}, and the
\emph{Thesaurus for the Social Sciences}, and
(3) DDI-RDF data sets provided by the \emph{Microdata Information System}, 
the \emph{Data Without Boundaries Discovery Portal}, the
\emph{Danish Data Archive}, and the
\emph{Swedish National Data Service}. We published the evaluation results for each QB data set in form of one document per SPARQL endpoint.\footnote{Online available at: \url{https://github.com/boschthomas/rdf-validation/tree/master/evaluation/data-sets/data-cube}}

\begin{table}[H]
		\scriptsize
    \begin{center}
		\caption{Validated Data Sets for each Vocabulary}
		\label{tab:datasets}
    \begin{tabular}{lrr}
           \textbf{Vocabulary}
           & \textbf{Data Sets}
           & \textbf{Triples}
					 
    \\ \midrule
		\textbf{QB} & $9,990$   & $3,775,983,610$  \\
		\textbf{SKOS} & $4,178$  & $477,737,281$ \\
		\textbf{DDI-RDF} & $1,526$  & $9,673,055$  \\
    \bottomrule
    \end{tabular}
    \end{center}
\end{table}

Since the validation of each of the 81 constraint types can be implemented using SPARQL, we use \emph{SPIN}, a SPARQL-based way to formulate and check constraints, as basis to develop a
validation environment to validate RDF data according to constraints expressed by arbitrary constraint languages\footnote{Constraint language implementations online available at: \url{https://github.com/boschthomas/rdf-validation/tree/master/SPIN}} \cite{BoschEckert2014-2}.
The \emph{RDF Validator}\footnote{Online demo available at: \url{http://purl.org/net/rdfval-demo}, source code online available at: \url{https://github.com/boschthomas/rdf-validator}} can directly be used to validate arbitrary RDF data for the three vocabularies. Additionally, own constraints on any vocabulary can be defined using several constraint languages.
The SPIN engine checks for each resource if it satisfies all constraints, which are associated with its assigned classes, and generates a result RDF graph containing information about all constraint violations.
There is one SPIN construct template for each constraint type.
A SPIN construct template contains a SPARQL CONSTRUCT query which generates constraint violation triples indicating the subject and the properties causing constraint violations and the reason why constraint violations have been raised.
A SPIN construct template creates constraint violation triples if all triple patterns within the SPARQL WHERE clause match.

\subsection{Evaluation Results}

Tables \ref{tab:evaluation-constraint-violations-1} and \ref{tab:evaluation-constraint-violations-2} show the results of the evaluation, more specifically the constraints and the constraint violations, which are caused by these constraints, in percent; whereas the numbers in the first line indicate the absolute amount of constraints and violations. The constraints and their raised violations are grouped by vocabulary, which type of language the constraint types are formulated with, and their severity level.
The numbers of validated triples and data sets differ between the vocabularies
as we validated 3.8 billion QB, 480 million SKOS, and 10 million DDI-RDF triples.
To be able to formulate findings which apply for all vocabularies, 
we only use normalized relative values representing the percentage of constraints and violations belonging to the respective sets. 

There is a strong overlap between \emph{RDFS/OWL} and \emph{Constraint Language Based} constraint types as in many cases constraint types are expressible by RDFS/OWL and classical constraint languages. This is the reason why the percentage values of constraints and violations grouped by the classification of constraint types according to the expressivity of constraint languages do not accumulate to 100\%. 

\begin{table}[H]
		\scriptsize
    \begin{center}
		\caption{Constraints and Constraint Violations (1)}
		\label{tab:evaluation-constraint-violations-1}
    \begin{tabular}{@{}lcc|cc@{}}
    \multirow{2}{*}{} &
      \multicolumn{2}{c}{\textbf{DDI-RDF}} &
      \multicolumn{2}{c}{\textbf{QB}} \\
    \textbf{} & C & CV & C & CV \\
    \hline
		 & 78 & 3,575,002 & 20 & 45,635,861 \\
		\hline
		\textbf{\emph{SPARQL}} & 29.5 & 34.7 & \textbf{60.0} & \textbf{100.0} \\
		\textbf{\emph{CL}} & \textbf{64.1} & \textbf{65.3} & 40.0 & 0.0 \\
		\textbf{\emph{RDFS/OWL}} & \textbf{66.7} & \textbf{65.3} & 40.0 & 0.0 \\
		\hline
		\textbf{\emph{info}} & \textbf{56.4} & \textbf{52.6} & 0.0 & 0.0 \\
		\textbf{\emph{warning}} & 11.5 & 29.4 & 15.0 & \textbf{99.8} \\
		\textbf{\emph{error}} & 32.1 & 18.0 & \textbf{85.0} & 0.3 \\
    \bottomrule
    \end{tabular}
    \\ \emph{C (constraints), CV (constraint violations)}
    \end{center}
\end{table}

\begin{table}[H]
		\scriptsize
    \begin{center}
		\caption{Constraints and Constraint Violations (2)}
		\label{tab:evaluation-constraint-violations-2}
    \begin{tabular}{@{}lcc|cc@{}}
    \multirow{2}{*}{} &
      \multicolumn{2}{c}{\textbf{SKOS}} &
      \multicolumn{2}{c}{\textbf{\emph{Total}}} \\
    \textbf{} & C & CV & C & CV \\
    \hline
		 & 17 & 5,540,988 & 115 & 54,751,851 \\
		\hline
		\textbf{\emph{SPARQL}} & \textbf{100.0} & \textbf{100.0} & \textbf{63.2} & \textbf{78.2} \\
		\textbf{\emph{CL}} & 0.0 & 0.0 & 34.7 & 21.8 \\
		\textbf{\emph{RDFS/OWL}} & 0.0 & 0.0 & 35.6 & 21.8 \\
		\hline
		\textbf{\emph{info}} & \textbf{70.6} & 41.2 & \textbf{42.3} & 31.3 \\
		\textbf{\emph{warning}} & 29.4 & \textbf{58.8} & 18.7 & \textbf{62.7} \\
		\textbf{\emph{error}} & 0.0 & 0.0 & \textbf{39.0} & 6.1\\
    \bottomrule
    \end{tabular}
    \\ \emph{C (constraints), CV (constraint violations)}
    \end{center}
\end{table}


\subsection{Legend}

In this sub-section, we describe how the tables in this paper should be read.
Table \ref{tab:legend} gives an overview over the symbols used in subsequent tables of the detailed evaluation.

\begin{table}[H]
	\centering
		\begin{tabular}{c|l}
      \textbf{Symbol} & \textbf{Description} \\	
			\hline
			\textbf{$\checkmark$} & Validation Successful (without any constraint violation) \\
			\textbf{\emph{X}} & Constraint Violations \\
			\hline
			\textbf{\textgreater \emph{X}} & Poor Performance/Scaling \\
      \ding{55} & Very Poor Performance/Scaling \\
			\hline
			\textbf{\emph{(!)}} & Not Yet Implemented Constraint \\
			\hline
			\textbf{\emph{(X)}} & The validation of \emph{X} data sets could not be finished, \\
			           & due to SPARQL endpoints' technical restrictions (e.g., defined timeouts). \\
			\hline
			\textbf{\emph{\textsuperscript{*}}} & default severity level \emph{informational} \\
			\textbf{\emph{\textsuperscript{**}}} & default severity level \emph{warning} \\
			\textbf{\emph{\textsuperscript{***}}} & default severity level \emph{error} \\
		\end{tabular}
	\caption{Legend}
	\label{tab:legend}
\end{table}

\begin{itemize}
	\item \textbf{Constraint Violations.}
When constraints are violated, 
\emph{X} indicates the number of raised constraint violation triples. 

  \item
\textbf{Poor Performance/Scaling.}
The performance of the implementation of the underlying SPARQL CONSTRUCT query 
is too poor to get all resulting constraint violation triples. 
Therefore, a limit of \emph{X} result constraint violation triples is set. 
It is likely that there are more than \emph{X} constraint violations.
Although the result set contains not the whole set of raised constraint violation triples,
the constraint can be used as an indicator if there is data not conforming to the constraint and
to resolve constraint violations step by step. 
As part of future work, the performance will be improved.

  \item
\textbf{Very Poor Performance/Scaling.} 
The performance of the implementation of the underlying SPARQL CONSTRUCT query 
is too poor to get any results, even though a limit of result constraint violation triples is set. 
As part of future work, the performance will be improved. 
\end{itemize}

\section{Evaluation of Metadata on Unit-Record Data Sets (DDI-RDF)}

In this section, the quality of the metadata on unit-record data sets (DDI-RDF) is evaluated by validating appropriate RDF constraints assigned to several RDF constraint types.
First, we give an overview on the evaluated data sets and finally we provide details about the evaluation.

\subsection{Data Sets Overview}

Tables \ref{tab:disco-data-sets-abbreviations} and \ref{tab:disco-sparql-endpoints} give an overview on the evaluated DDI-RDF data sets, their abbreviations, and publicly available SPARQL endpoints.
Table \ref{tab:disco-overview} comprehends the number of triples, data sets, and instances of multiple vocabulary-specific classes.

\begin{table}[H]
	\centering
		\begin{tabular}{l|l}
      \textbf{Abbr.} & \textbf{DDI-RDF Data Sets} \\		
      \hline
    \emph{Missy} & \emph{Microdata Information System}\tablefootnote{\url{http://www.gesis.org/missy/eu/missy-home}} \\
		\emph{DwB} & \emph{DwB Discovery Portal}\tablefootnote{\url{http://dwb-dev.nsd.uib.no/portal}} \\
		\emph{DDA-SND} & \emph{DDI-RDF}\tablefootnote{\url{http://ddi-rdf.borsna.se/}} \\
		               & provided by the \emph{Danish Data Archive (DDA)}\tablefootnote{\url{http://samfund.dda.dk/dda/default-en.asp}} and Swedish National Data Service (SND)\tablefootnote{\url{http://snd.gu.se/en}} \\ 
		\end{tabular}
	\caption{DDI-RDF Data Sets Abbreviations}
	\label{tab:disco-data-sets-abbreviations}
\end{table}

\begin{table}[H]
    \begin{center}
    \begin{tabular}{@{}lccccccccccc@{}}
           & \multicolumn{9}{c}{\textbf{Counts}}
    \\  \cmidrule{2-10}
    \\       \textbf{Data Sets}
           & \textbf{\rot{triples}}
           & \textbf{\rot{disco:StudyGroup}}
           & \textbf{\rot{disco:Study}}
           & \textbf{\rot{disco:LogicalDataSet}}
           & \textbf{\rot{disco:Universe}}
					 & \textbf{\rot{disco:Variable}}
					 & \textbf{\rot{disco:Question}}
				   & \textbf{\rot{disco:SummaryStatistics}}
					 & \textbf{\rot{disco:CategoryStatistics}}
					 & \textbf{\rot{skos:Concept}}
    \\ \midrule
    \emph{Missy} & 5,068,838 & 6 & 45 & 159 & 1,125 & 21,040 & 0 & 0 & 0 & 147,193 \\
		\emph{DwB} & 2,332,802 & 0 & 1,387 & 1,367 & 2,796 & 446,806 & 0 & 0 & 0 & 0 \\
		\emph{DDA-SND} & 2,271,415 & 0 & 1,490 & 0 & 10,188 & 80,070 & 139,237 & 0 & 0 & 290,963 \\ 
		\hline
		\textbf{\emph{Total}} & 9,673,055 & & & 1,526 \\
    \bottomrule
    \end{tabular}
    \caption{DDI-RDF Data Sets Overview}
		\label{tab:disco-overview}
    \end{center}
\end{table}

\begin{table}[H]
	\centering
		\begin{tabular}{l|l}
      \textbf{Data Sets} & \textbf{SPARQL Endpoint} \\		
      \hline
      \emph{Missy} & \url{http://svko-missy:8181/openrdf-workbench/repositories/native-java-store/summary} \\
			\emph{DwB} & \url{http://dwb-dev.nsd.uib.no/sparql} \\
			\emph{DDA-SND} & \url{http://ddi-rdf.borsna.se/endpoint/} \\
		\end{tabular}
	\caption{DDI-RDF SPARQL Endpoints}
	\label{tab:disco-sparql-endpoints}
\end{table}

\subsection{Detailed Evaluation}

In this sub-section, we give details about the evaluation in form of diverse tables containing the number of constraint violations per evaluated data set and constraint of particular constraint types.

\begin{table}[H]
    \begin{center}
    \begin{tabular}{@{}lccc@{}}
           & \multicolumn{3}{c}{\textbf{Data Sets}}
    \\  \cmidrule{2-4}
    \\       \textbf{Existential Quantifications (1)}
           & \rot{\emph{Missy}}
           & \rot{\emph{DwB}}
           & \rot{\emph{DDA-SND}}
    \\ \midrule
    \emph{DISCO-C-EXISTENTIAL-QUANTIFICATIONS-01}\textsuperscript{***} & $\checkmark$ & $\checkmark$ & $\checkmark$ \\
		\emph{DISCO-C-EXISTENTIAL-QUANTIFICATIONS-02}\textsuperscript{***} & 7 & 17 & 1,490 \\
		\emph{DISCO-C-EXISTENTIAL-QUANTIFICATIONS-03}\textsuperscript{*} & $\checkmark$ & $\checkmark$ & $\checkmark$ \\
		\emph{DISCO-C-EXISTENTIAL-QUANTIFICATIONS-04}\textsuperscript{*} & 11,021 & 445,381 & 62,260 \\
		\emph{DISCO-C-EXISTENTIAL-QUANTIFICATIONS-05}\textsuperscript{*} & $\checkmark$ & $\checkmark$ & 139,237 \\
		\emph{DISCO-C-EXISTENTIAL-QUANTIFICATIONS-06}\textsuperscript{*} & 12 & 1,367 & $\checkmark$ \\
		\emph{DISCO-C-EXISTENTIAL-QUANTIFICATIONS-07}\textsuperscript{*} & 6 & $\checkmark$ & $\checkmark$ \\
		\emph{DISCO-C-EXISTENTIAL-QUANTIFICATIONS-08}\textsuperscript{*} & 45 & 1,387 & 1,490 \\
		\emph{DISCO-C-EXISTENTIAL-QUANTIFICATIONS-09}\textsuperscript{*} & 6 & $\checkmark$ & $\checkmark$ \\
		\emph{DISCO-C-EXISTENTIAL-QUANTIFICATIONS-10}\textsuperscript{*} & 45 & 1,387 & 1,490 \\
    \bottomrule
    \end{tabular}
    \caption{Evaluation of DDI-RDF Data Sets - Existential Quantifications (1)}
		\label{tab:evaluation-disco-existential-quantifications-1}
    \end{center}
\end{table}

\begin{table}[H]
    \begin{center}
    \begin{tabular}{@{}lccc@{}}
           & \multicolumn{3}{c}{\textbf{Data Sets}}
    \\  \cmidrule{2-4}
    \\       \textbf{Existential Quantifications (2)}
           & \rot{\emph{Missy}}
           & \rot{\emph{DwB}}
           & \rot{\emph{DDA-SND}}
    \\ \midrule
    \emph{DISCO-C-EXISTENTIAL-QUANTIFICATIONS-11}\textsuperscript{*} & 6 & $\checkmark$ & $\checkmark$ \\
		\emph{DISCO-C-EXISTENTIAL-QUANTIFICATIONS-12}\textsuperscript{*} & 6 & $\checkmark$ & $\checkmark$ \\
		\emph{DISCO-C-EXISTENTIAL-QUANTIFICATIONS-13}\textsuperscript{*} & $\checkmark$ & $\checkmark$ & $\checkmark$ \\
		\emph{DISCO-C-EXISTENTIAL-QUANTIFICATIONS-14}\textsuperscript{*} & 45 & 1,387 & 1,490 \\
		\emph{DISCO-C-EXISTENTIAL-QUANTIFICATIONS-15}\textsuperscript{*} & 45 & 1,387 & 1,490 \\
		\emph{DISCO-C-EXISTENTIAL-QUANTIFICATIONS-16}\textsuperscript{*} & $\checkmark$ & $\checkmark$ & $\checkmark$ \\
		\emph{DISCO-C-EXISTENTIAL-QUANTIFICATIONS-17}\textsuperscript{*} & 159 & 1,367 & $\checkmark$ \\
		\emph{DISCO-C-EXISTENTIAL-QUANTIFICATIONS-18}\textsuperscript{*} & 159 & 1,367 & $\checkmark$ \\
		\emph{DISCO-C-EXISTENTIAL-QUANTIFICATIONS-19}\textsuperscript{*} & $\checkmark$ & $\checkmark$ & $\checkmark$ \\
		\emph{DISCO-C-EXISTENTIAL-QUANTIFICATIONS-20}\textsuperscript{*} & $\checkmark$ & 1,367 & $\checkmark$ \\
    \bottomrule
    \end{tabular}
    \caption{Evaluation of DDI-RDF Data Sets - Existential Quantifications (2)}
		\label{tab:evaluation-disco-existential-quantifications-2}
    \end{center}
\end{table}

\begin{table}[H]
    \begin{center}
    \begin{tabular}{@{}lccc@{}}
           & \multicolumn{3}{c}{\textbf{Data Sets}}
    \\  \cmidrule{2-4}
    \\       \textbf{Existential Quantifications (3)}
           & \rot{\emph{Missy}}
           & \rot{\emph{DwB}}
           & \rot{\emph{DDA-SND}}
    \\ \midrule
    \emph{DISCO-C-EXISTENTIAL-QUANTIFICATIONS-21}\textsuperscript{*} & $\checkmark$ & 1,367 & $\checkmark$ \\
		\emph{DISCO-C-EXISTENTIAL-QUANTIFICATIONS-22}\textsuperscript{*} & $\checkmark$ & $\checkmark$ & $\checkmark$ \\
		\emph{DISCO-C-EXISTENTIAL-QUANTIFICATIONS-23}\textsuperscript{*} & 6 & $\checkmark$ & $\checkmark$ \\
		\emph{DISCO-C-EXISTENTIAL-QUANTIFICATIONS-24}\textsuperscript{*} & 45 & 1,387 & 1,490 \\
		\emph{DISCO-C-EXISTENTIAL-QUANTIFICATIONS-25}\textsuperscript{*} & 45 & 1,387 & 1,490 \\
		\emph{DISCO-C-EXISTENTIAL-QUANTIFICATIONS-26}\textsuperscript{*} & 45 & 1,387 & 1,490 \\
		\emph{DISCO-C-EXISTENTIAL-QUANTIFICATIONS-27}\textsuperscript{***} & $\checkmark$ & 130 & 1,490 \\
		\emph{DISCO-C-EXISTENTIAL-QUANTIFICATIONS-28}\textsuperscript{**} & 159 & $\checkmark$ & $\checkmark$ \\
		\emph{DISCO-C-EXISTENTIAL-QUANTIFICATIONS-29}\textsuperscript{**} & $\checkmark$ & $\checkmark$ & $\checkmark$ \\
		\emph{DISCO-C-EXISTENTIAL-QUANTIFICATIONS-30}\textsuperscript{**} & $\checkmark$ & $\checkmark$ & $\checkmark$ \\
    \bottomrule
    \end{tabular}
    \caption{Evaluation of DDI-RDF Data Sets - Existential Quantifications (3)}
		\label{tab:evaluation-disco-existential-quantifications-3}
    \end{center}
\end{table}

\begin{table}[H]
    \begin{center}
    \begin{tabular}{@{}lccc@{}}
           & \multicolumn{3}{c}{\textbf{Data Sets}}
    \\  \cmidrule{2-4}
    \\       \textbf{Existential Quantifications (4)}
           & \rot{\emph{Missy}}
           & \rot{\emph{DwB}}
           & \rot{\emph{DDA-SND}}
    \\ \midrule
		\emph{DISCO-C-EXISTENTIAL-QUANTIFICATIONS-31}\textsuperscript{**} & 159 & 1,367 & $\checkmark$ \\
		\emph{DISCO-C-EXISTENTIAL-QUANTIFICATIONS-32}\textsuperscript{***} & $\checkmark$ & $\checkmark$ & $\checkmark$ \\
		\emph{DISCO-C-EXISTENTIAL-QUANTIFICATIONS-33}\textsuperscript{***} & $\checkmark$ & $\checkmark$ & $\checkmark$ \\
		\emph{DISCO-C-EXISTENTIAL-QUANTIFICATIONS-34}\textsuperscript{***} & $\checkmark$ & $\checkmark$ & $\checkmark$ \\
		\emph{DISCO-C-EXISTENTIAL-QUANTIFICATIONS-35}\textsuperscript{***} & $\checkmark$ & $\checkmark$ & $\checkmark$ \\
		\emph{DISCO-C-EXISTENTIAL-QUANTIFICATIONS-36}\textsuperscript{***} & $\checkmark$ & $\checkmark$ & $\checkmark$ \\
		\emph{DISCO-C-EXISTENTIAL-QUANTIFICATIONS-37}\textsuperscript{*} & 18,625 & $\checkmark$ & $\checkmark$ \\
		\emph{DISCO-C-EXISTENTIAL-QUANTIFICATIONS-38}\textsuperscript{*} & $\checkmark$ & $\checkmark$ & 750 \\
		\emph{DISCO-C-EXISTENTIAL-QUANTIFICATIONS-39}\textsuperscript{***} & $\checkmark$ & $\checkmark$ & $\checkmark$ \\
		\emph{DISCO-C-EXISTENTIAL-QUANTIFICATIONS-40}\textsuperscript{*} & $\checkmark$ & $\checkmark$ & 139,237 \\
    \bottomrule
    \end{tabular}
    \caption{Evaluation of DDI-RDF Data Sets - Existential Quantifications (4)}
		\label{tab:evaluation-disco-existential-quantifications-4}
    \end{center}
\end{table}

\begin{table}[H]
    \begin{center}
    \begin{tabular}{@{}lccc@{}}
           & \multicolumn{3}{c}{\textbf{Data Sets}}
    \\  \cmidrule{2-4}
    \\       \textbf{Existential Quantifications (5)}
           & \rot{\emph{Missy}}
           & \rot{\emph{DwB}}
           & \rot{\emph{DDA-SND}}
    \\ \midrule
		\emph{DISCO-C-EXISTENTIAL-QUANTIFICATIONS-41}\textsuperscript{*} & $\checkmark$ & $\checkmark$ & $\checkmark$ \\
		\emph{DISCO-C-EXISTENTIAL-QUANTIFICATIONS-42}\textsuperscript{*} & $\checkmark$ & $\checkmark$ & $\checkmark$ \\
		\emph{DISCO-C-EXISTENTIAL-QUANTIFICATIONS-43}\textsuperscript{*} & 15,733 & 446,806 & 80,070 \\
		\emph{DISCO-C-EXISTENTIAL-QUANTIFICATIONS-44}\textsuperscript{*} & 159 & $\checkmark$ & $\checkmark$ \\
		\emph{DISCO-C-EXISTENTIAL-QUANTIFICATIONS-45}\textsuperscript{*} & 6,784 & 446,806 & 19,221 \\
		\emph{DISCO-C-EXISTENTIAL-QUANTIFICATIONS-46}\textsuperscript{**} & 11,550 & 446,806 & 10,451 \\
    \bottomrule
    \end{tabular}
    \caption{Evaluation of DDI-RDF Data Sets - Existential Quantifications (5)}
		\label{tab:evaluation-disco-existential-quantifications-5}
    \end{center}
\end{table}

\begin{table}[H]
    \begin{center}
    \begin{tabular}{@{}lccc@{}}
           & \multicolumn{3}{c}{\textbf{Data Sets}}
    \\  \cmidrule{2-4}
    \\       \textbf{Conditional Properties}
           & \rot{\emph{Missy}}
           & \rot{\emph{DwB}}
           & \rot{\emph{DDA-SND}}
    \\ \midrule
    \emph{ DISCO-C-CONDITIONAL-PROPERTIES-01}\textsuperscript{***} & $\checkmark$ & $\checkmark$ & 80,070 \\
		\emph{ DISCO-C-CONDITIONAL-PROPERTIES-02}\textsuperscript{**} & 12 & $\checkmark$ & $\checkmark$ \\
		\emph{ DISCO-C-CONDITIONAL-PROPERTIES-03}\textsuperscript{**} & 90 & $\checkmark$ & 2,980 \\
		\emph{ DISCO-C-CONDITIONAL-PROPERTIES-04}\textsuperscript{***} & 6 & $\checkmark$ & $\checkmark$ \\
		\emph{ DISCO-C-CONDITIONAL-PROPERTIES-05}\textsuperscript{***} & 45 & 1,387 & 1,490 \\
		\emph{ DISCO-C-CONDITIONAL-PROPERTIES-06}\textsuperscript{***} & $\checkmark$ & $\checkmark$ & $\checkmark$ \\
    \bottomrule
    \end{tabular}
    \caption{Evaluation of DDI-RDF Data Sets - Conditional Properties}
		\label{tab:evaluation-disco-conditional-properties}
    \end{center}
\end{table}

\begin{table}[H]
    \begin{center}
    \begin{tabular}{@{}lccc@{}}
           & \multicolumn{3}{c}{\textbf{Data Sets}}
    \\  \cmidrule{2-4}
    \\       \textbf{Provenance}
           & \rot{\emph{Missy}}
           & \rot{\emph{DwB}}
           & \rot{\emph{DDA-SND}}
    \\ \midrule
    \emph{DISCO-C-PROVENANCE-01}\textsuperscript{*} & 6 & $\checkmark$ & $\checkmark$ \\
		\emph{DISCO-C-PROVENANCE-02}\textsuperscript{*} & 45 & 1,387 & 1,490 \\
		\emph{DISCO-C-PROVENANCE-03}\textsuperscript{*} & 159 & 1,367 & $\checkmark$ \\
		\emph{DISCO-C-PROVENANCE-04}\textsuperscript{*} & $\checkmark$ & 1,367 & $\checkmark$ \\
    \bottomrule
    \end{tabular}
    \caption{Evaluation of DDI-RDF Data Sets - Provenance}
		\label{tab:evaluation-disco-provenance}
    \end{center}
\end{table}

\begin{table}[H]
    \begin{center}
    \begin{tabular}{@{}lccc@{}}
           & \multicolumn{3}{c}{\textbf{Data Sets}}
    \\  \cmidrule{2-4}
    \\       \textbf{Labeling and Documentation}
           & \rot{\emph{Missy}}
           & \rot{\emph{DwB}}
           & \rot{\emph{DDA-SND}}
    \\ \midrule
    \emph{DISCO-C-LABELING-AND-DOCUMENTATION-01}\textsuperscript{*} & 6 & $\checkmark$ & $\checkmark$ \\
		\emph{DISCO-C-LABELING-AND-DOCUMENTATION-02}\textsuperscript{*} & 45 & 1,387 & 1,490 \\
		\emph{DISCO-C-LABELING-AND-DOCUMENTATION-03}\textsuperscript{*} & 159 & 1,367 & $\checkmark$ \\
		\emph{DISCO-C-LABELING-AND-DOCUMENTATION-04}\textsuperscript{*} & $\checkmark$ & 1,367 & $\checkmark$ \\
		\emph{DISCO-C-LABELING-AND-DOCUMENTATION-05}\textsuperscript{*} & $\checkmark$ & $\checkmark$ & $\checkmark$ \\
		\emph{DISCO-C-LABELING-AND-DOCUMENTATION-06}\textsuperscript{*} & 21,040 & 446,806 & 80,070 \\
    \bottomrule
    \end{tabular}
    \caption{Evaluation of DDI-RDF Data Sets - Labeling and Documentation}
		\label{tab:evaluation-disco-labeling-and-documentation}
    \end{center}
\end{table}

\begin{table}[H]
    \begin{center}
    \begin{tabular}{@{}lccc@{}}
           & \multicolumn{3}{c}{\textbf{Data Sets}}
    \\  \cmidrule{2-4}
    \\       \textbf{Data Model Consistency}
           & \rot{\emph{Missy}}
           & \rot{\emph{DwB}}
           & \rot{\emph{DDA-SND}}
    \\ \midrule
    \emph{DISCO-C-DATA-MODEL-CONSISTENCY-01 (!)}\textsuperscript{***} \\
		\emph{DISCO-C-DATA-MODEL-CONSISTENCY-02 (!)}\textsuperscript{***} \\
		\emph{DISCO-C-DATA-MODEL-CONSISTENCY-03 (!)}\textsuperscript{***} \\
		\emph{DISCO-C-DATA-MODEL-CONSISTENCY-04 (!)}\textsuperscript{***} \\
		\emph{DISCO-C-DATA-MODEL-CONSISTENCY-05}\textsuperscript{***} & $\checkmark$ & $\checkmark$ & $\checkmark$ \\
		\emph{DISCO-C-DATA-MODEL-CONSISTENCY-06 (!)}\textsuperscript{***} \\
		\emph{DISCO-C-DATA-MODEL-CONSISTENCY-07 (!)}\textsuperscript{***} \\
    \bottomrule
    \end{tabular}
    \caption{Evaluation of DDI-RDF Data Sets - Data Model Consistency}
		\label{tab:evaluation-disco-data-model-consistency}
    \end{center}
\end{table}

\begin{table}[H]
    \begin{center}
    \begin{tabular}{@{}lccc@{}}
           & \multicolumn{3}{c}{\textbf{Data Sets}}
    \\  \cmidrule{2-4}
    \\       \textbf{Comparison}
           & \rot{\emph{Missy}}
           & \rot{\emph{DwB}}
           & \rot{\emph{DDA-SND}}
    \\ \midrule
    \emph{DISCO-C-COMPARISON-VARIABLES-01 (!)}\textsuperscript{**} \\
		\emph{DISCO-C-COMPARISON-VARIABLES-02}\textsuperscript{***} & 21,040 & 446,806 & 80,070 \\
		\emph{DISCO-C-COMPARISON-VARIABLES-03 (!)}\textsuperscript{***} \\
		\emph{DISCO-C-COMPARISON-VARIABLES-04}\textsuperscript{*} & 18,625 & $\checkmark$ & $\checkmark$ \\
		\emph{DISCO-C-COMPARISON-VARIABLES-05}\textsuperscript{***} & 159 & $\checkmark$ & $\checkmark$ \\
    \bottomrule
    \end{tabular}
    \caption{Evaluation of DDI-RDF Data Sets - Comparison}
		\label{tab:evaluation-disco-comparison}
    \end{center}
\end{table}

\begin{table}[H]
    \begin{center}
    \begin{tabular}{@{}lccc@{}}
           & \multicolumn{3}{c}{\textbf{Data Sets}}
    \\  \cmidrule{2-4}
    \\       \textbf{Mathematical Operations}
           & \rot{\emph{Missy}}
           & \rot{\emph{DwB}}
           & \rot{\emph{DDA-SND}}
    \\ \midrule
    \emph{DISCO-C-MATHEMATICAL-OPERATIONS-01 (!)}\textsuperscript{***} \\
		\emph{DISCO-C-MATHEMATICAL-OPERATIONS-02 (!)}\textsuperscript{***} \\
		\emph{DISCO-C-MATHEMATICAL-OPERATIONS-03 (!)}\textsuperscript{***} \\
		\emph{DISCO-C-MATHEMATICAL-OPERATIONS-04 (!)}\textsuperscript{***} \\
		\emph{DISCO-C-MATHEMATICAL-OPERATIONS-05 (!)}\textsuperscript{***} \\
    \bottomrule
    \end{tabular}
    \caption{Evaluation of DDI-RDF Data Sets - Mathematical Operations}
		\label{tab:evaluation-disco-mathematical-operations}
    \end{center}
\end{table}

\begin{table}[H]
    \begin{center}
    \begin{tabular}{@{}lccc@{}}
           & \multicolumn{3}{c}{\textbf{Data Sets}}
    \\  \cmidrule{2-4}
    \\       \textbf{Language Tags}
           & \rot{\emph{Missy}}
           & \rot{\emph{DwB}}
           & \rot{\emph{DDA-SND}}
    \\ \midrule
    \emph{DISCO-C-LANGUAGE-TAG-MATCHING-01 (!)}\textsuperscript{*} \\
		\emph{DISCO-C-LANGUAGE-TAG-CARDINALITY-01 (!)}\textsuperscript{*} \\
		\emph{DISCO-C-LANGUAGE-TAG-CARDINALITY-02 (!)}\textsuperscript{*} \\
		\emph{DISCO-C-LANGUAGE-TAG-CARDINALITY-03 (!)}\textsuperscript{*} \\
    \bottomrule
    \end{tabular}
    \caption{Evaluation of DDI-RDF Data Sets - Language Tags}
		\label{tab:evaluation-disco-language-tags}
    \end{center}
\end{table}

\begin{table}[H]
    \begin{center}
    \begin{tabular}{@{}lccc@{}}
           & \multicolumn{3}{c}{\textbf{Data Sets}}
    \\  \cmidrule{2-4}
    \\       \textbf{Aggregation}
           & \rot{\emph{Missy}}
           & \rot{\emph{DwB}}
           & \rot{\emph{DDA-SND}}
    \\ \midrule
    \emph{DISCO-C-AGGREGATION-01 (!)}\textsuperscript{*} \\
		\emph{DISCO-C-AGGREGATION-02 (!)}\textsuperscript{*} \\
		\emph{DISCO-C-AGGREGATION-03 (!)}\textsuperscript{*} \\
		\emph{DISCO-C-AGGREGATION-04 (!)}\textsuperscript{*} \\
		\emph{DISCO-C-AGGREGATION-05 (!)}\textsuperscript{*} \\
		\emph{DISCO-C-AGGREGATION-06 (!)}\textsuperscript{*} \\
		\emph{DISCO-C-AGGREGATION-07 (!)}\textsuperscript{*} \\
    \bottomrule
    \end{tabular}
    \caption{Evaluation of DDI-RDF Data Sets - Aggregation}
		\label{tab:evaluation-disco-aggregation}
    \end{center}
\end{table}

\begin{table}[H]
    \begin{center}
    \begin{tabular}{@{}lccc@{}}
           & \multicolumn{3}{c}{\textbf{Data Sets}}
    \\  \cmidrule{2-4}
    \\       \textbf{DDI-RDF Constraints}
           & \rot{\emph{Missy}}
           & \rot{\emph{DwB}}
           & \rot{\emph{DDA-SND}}
    \\ \midrule
    \emph{DISCO-C-ALLOWED-VALUES-01}\textsuperscript{***} & $\checkmark$ & $\checkmark$ & $\checkmark$ \\
		\emph{DISCO-C-LITERAL-RANGES-01}\textsuperscript{***} & $\checkmark$ & $\checkmark$ & $\checkmark$ \\
		\emph{DISCO-C-INVERSE-FUNCTIONAL-PROPERTIES-01}\textsuperscript{***} & $\checkmark$ & $\checkmark$ & $\checkmark$ \\
		\emph{DISCO-C-INVERSE-FUNCTIONAL-PROPERTIES-02}\textsuperscript{***} & $\checkmark$ & $\checkmark$ & $\checkmark$ \\
		\emph{DISCO-C-CLASS-SPECIFIC-PROPERTY-RANGE-01}\textsuperscript{***} & $\checkmark$ & $\checkmark$ & $\checkmark$ \\
		\emph{DISCO-C-MEMBERSHIP-IN-CONTROLLED-VOCABULARIES-01}\textsuperscript{***} & $\checkmark$ & $\checkmark$ & \ding{55} \\
		\emph{DISCO-C-LITERAL-VALUE-COMPARISON-01}\textsuperscript{***} & $\checkmark$ & 1,299 & $\checkmark$ \\
		\emph{DISCO-C-CONTEXT-SPECIFIC-VALID-PROPERTIES-01}\textsuperscript{*} & 21,038 & $\checkmark$ & $\checkmark$ \\
		\emph{DISCO-C-DATA-PROPERTY-FACETS-01}\textsuperscript{**} & $\checkmark$ & $\checkmark$ & $\checkmark$ \\
		\emph{DISCO-C-DATA-PROPERTY-FACETS-02}\textsuperscript{**} & $\checkmark$ & $\checkmark$ & $\checkmark$ \\
    \bottomrule
    \end{tabular}
    \caption{Evaluation of DDI-RDF Data Sets - DDI-RDF Constraints (1)}
		\label{tab:evaluation-disco-disco-constraints-1}
    \end{center}
\end{table}

\begin{table}[H]
    \begin{center}
    \begin{tabular}{@{}lccc@{}}
           & \multicolumn{3}{c}{\textbf{Data Sets}}
    \\  \cmidrule{2-4}
    \\       \textbf{DDI-RDF Constraints}
           & \rot{\emph{Missy}}
           & \rot{\emph{DwB}}
           & \rot{\emph{DDA-SND}}
    \\ \midrule
		\emph{DISCO-C-VALUE-IS-VALID-FOR-DATATYPE-01}\textsuperscript{***} & 30 & 6,932 & $\checkmark$ \\
		\emph{DISCO-C-VALUE-IS-VALID-FOR-DATATYPE-02}\textsuperscript{***} & $\checkmark$ & $\checkmark$ & $\checkmark$ \\
		\emph{DISCO-C-SUBSUMPTION-01 (!)}\textsuperscript{***}\textsuperscript{B} \\
		\emph{DISCO-C-CLASS-EQUIVALENCE-01 (!)}\textsuperscript{*} \\
		\emph{DISCO-C-SUB-PROPERTIES-01 (!)}\textsuperscript{***} \\
		\emph{DISCO-C-PROPERTY-DOMAIN-01 (!)}\textsuperscript{***} \\
		\emph{DISCO-C-PROPERTY-RANGES-01 (!)}\textsuperscript{***} \\
		\emph{DISCO-C-INVERSE-OBJECT-PROPERTIES-01 (!)}\textsuperscript{***} \\
		\emph{DISCO-C-INVERSE-OBJECT-PROPERTIES-02 (!)}\textsuperscript{***} \\
		\emph{DISCO-C-INVERSE-OBJECT-PROPERTIES-03 (!)}\textsuperscript{***} \\
		\emph{DISCO-C-DISJOINT-PROPERTIES-01 (!)}\textsuperscript{***} \\
    \bottomrule
    \end{tabular}
    \caption{Evaluation of DDI-RDF Data Sets - DDI-RDF Constraints (2)}
		\label{tab:evaluation-disco-disco-constraints-2}
    \end{center}
\end{table}

\begin{table}[H]
    \begin{center}
    \begin{tabular}{@{}lccc@{}}
           & \multicolumn{3}{c}{\textbf{Data Sets}}
    \\  \cmidrule{2-4}
    \\       \textbf{DDI-RDF Constraints}
           & \rot{\emph{Missy}}
           & \rot{\emph{DwB}}
           & \rot{\emph{DDA-SND}}
    \\ \midrule
		\emph{DISCO-C-ASYMMETRIC-OBJECT-PROPERTIES-01 (!)}\textsuperscript{***} \\
		\emph{DISCO-C-IRREFLEXIVE-OBJECT-PROPERTIES-01 (!)}\textsuperscript{***} \\
		\emph{DISCO-C-CLASS-SPECIFIC-IRREFLEXIVE-OBJECT-PROPERTIES-01 (!)}\textsuperscript{***} \\
		\emph{DISCO-C-CLASS-SPECIFIC-IRREFLEXIVE-OBJECT-PROPERTIES-02 (!)}\textsuperscript{***} \\
		\emph{DISCO-C-DISJOINT-CLASSES-01 (!)}\textsuperscript{***} \\
		\emph{DISCO-C-EQUIVALENT-PROPERTIES-01 (!)}\textsuperscript{*} \\
		\emph{DISCO-C-LITERAL-PATTERN-MATCHING-01 (!)}\textsuperscript{*} \\
		\emph{DISCO-C-DISJUNCTION-01 (!)}\textsuperscript{***} \\
		\emph{DISCO-C-UNIVERSAL-QUANTIFICATIONS-01 (!)}\textsuperscript{***} \\
		\emph{DISCO-C-MINIMUM-QUALIFIED-CARDINALITY-RESTRICTIONS-01 (!)}\textsuperscript{***} \\
    \bottomrule
    \end{tabular}
    \caption{Evaluation of DDI-RDF Data Sets - DDI-RDF Constraints (3)}
		\label{tab:evaluation-disco-disco-constraints-3}
    \end{center}
\end{table}

\begin{table}[H]
    \begin{center}
    \begin{tabular}{@{}lccc@{}}
           & \multicolumn{3}{c}{\textbf{Data Sets}}
    \\  \cmidrule{2-4}
    \\       \textbf{DDI-RDF Constraints}
           & \rot{\emph{Missy}}
           & \rot{\emph{DwB}}
           & \rot{\emph{DDA-SND}}
    \\ \midrule
		\emph{DISCO-C-MAXIMUM-QUALIFIED-CARDINALITY-RESTRICTIONS-01 (!)}\textsuperscript{***} \\
		\emph{DISCO-C-EXACT-QUALIFIED-CARDINALITY-RESTRICTIONS-01 (!)}\textsuperscript{***} \\
		\emph{DISCO-C-CONTEXT-SPECIFIC-EXCLUSIVE-OR-OF-PROPERTY-GROUPS-01 (!)}\textsuperscript{*} \\
		\emph{DISCO-C-IRI-PATTERN-MATCHING-01 (!)}\textsuperscript{*} \\
		\emph{DISCO-C-ORDERING-01 (!)}\textsuperscript{*} \\
		\emph{DISCO-C-ORDERING-02 (!)}\textsuperscript{*} \\
		\emph{DISCO-C-ORDERING-03 (!)}\textsuperscript{*} \\
		\emph{DISCO-C-STRING-OPERATIONS-01 (!)}\textsuperscript{*} \\
		\emph{DISCO-C-CONTEXT-SPECIFIC-VALID-CLASSES-01 (!)}\textsuperscript{*} \\
		\emph{DISCO-C-CONTEXT-SPECIFIC-VALID-PROPERTIES-01 (!)}\textsuperscript{*} \\
    \bottomrule
    \end{tabular}
    \caption{Evaluation of DDI-RDF Data Sets - DDI-RDF Constraints (4)}
		\label{tab:evaluation-disco-disco-constraints-4}
    \end{center}
\end{table}

\begin{table}[H]
    \begin{center}
    \begin{tabular}{@{}lccc@{}}
           & \multicolumn{3}{c}{\textbf{Data Sets}}
    \\  \cmidrule{2-4}
    \\       \textbf{DDI-RDF Constraints}
           & \rot{\emph{Missy}}
           & \rot{\emph{DwB}}
           & \rot{\emph{DDA-SND}}
    \\ \midrule
		\emph{DISCO-C-DEFAULT-VALUES-01 (!)}\textsuperscript{*} \\
		\emph{DISCO-C-WHITESPACE-HANDLING-01 (!)}\textsuperscript{*} \\
		\emph{DISCO-C-HTML-HANDLING-01 (!)}\textsuperscript{*} \\
		\emph{DISCO-C-HTML-HANDLING-02 (!)}\textsuperscript{*} \\
		\emph{DISCO-C-RECOMMENDED-PROPERTIES-01 (!)}\textsuperscript{*} \\
		\emph{DISCO-C-HANDLE-RDF-COLLECTIONS-01 (!)}\textsuperscript{*} \\
		\emph{DISCO-C-HANDLE-RDF-COLLECTIONS-02 (!)}\textsuperscript{*} \\
		\emph{DISCO-C-USE-SUB-SUPER-RELATIONS-IN-VALIDATION-01 (!)}\textsuperscript{*} \\
		\emph{DISCO-C-USE-SUB-SUPER-RELATIONS-IN-VALIDATION-02 (!)}\textsuperscript{*} \\
		\emph{DISCO-C-STRUCTURE-01 (!)}\textsuperscript{***} \\
    \bottomrule
    \end{tabular}
    \caption{Evaluation of DDI-RDF Data Sets - DDI-RDF Constraints (5)}
		\label{tab:evaluation-disco-disco-constraints-5}
    \end{center}
\end{table}

\begin{table}[H]
    \begin{center}
    \begin{tabular}{@{}lccc@{}}
           & \multicolumn{3}{c}{\textbf{Data Sets}}
    \\  \cmidrule{2-4}
    \\       \textbf{DDI-RDF Constraints}
           & \rot{\emph{Missy}}
           & \rot{\emph{DwB}}
           & \rot{\emph{DDA-SND}}
    \\ \midrule
		\emph{DISCO-C-VOCABULARY-01 (!)}\textsuperscript{***} \\
		\emph{DISCO-C-HTTP-URI-SCHEME-VIOLATION (!)}\textsuperscript{***} \\
    \bottomrule
    \end{tabular}
    \caption{Evaluation of DDI-RDF Data Sets - DDI-RDF Constraints (6)}
		\label{tab:evaluation-disco-disco-constraints-6}
    \end{center}
\end{table}

\section{Evaluation of Metadata and Data of Aggregated Data Sets (QB)}

In this section, the quality of the metadata on aggregated data (QB) data sets and of the data sets themselves is evaluated by validating appropriate RDF constraints assigned to several RDF constraint types.
First, we we give an overview on the evaluated data sets and finally we provide details about the evaluation.

\subsection{Data Sets Overview}

There are websites giving an overview on available QB data sets\footnote{\url{http://270a.info/}; \url{http://datahub.io/de/dataset?tags=format-qb}; \url{http://ontologycentral.com/}}. 
Tables \ref{tab:data-cube-data-sets-abbreviations} and \ref{tab:data-cube-sparql-endpoints} give an overview on the evaluated QB data sets, their abbreviations, and publicly available SPARQL endpoints.
Table \ref{tab:data-cube-overview} comprehends the number of triples, data sets, and instances of multiple vocabulary-specific classes.


\begin{table}[H]
	\centering
		\begin{tabular}{l|l}
      \textbf{Abbr.} & \textbf{QB Data Sets} \\		
      \hline
    \emph{ECB} & \emph{European Central Bank}\tablefootnote{\url{http://www.ecb.europa.eu/home/html/index.en.html}} \\
		\emph{UIS} & \emph{UNESCO Institute for Statistics}\tablefootnote{\url{http://www.uis.unesco.org/Pages/default.aspx}} \\
		\emph{IMF} & \emph{International Monetary Fund}\tablefootnote{\url{http://www.imf.org/external/index.htm}} \\
		\emph{BFS} & \emph{Bundesamt für Statistik - Swiss Federal Statistics}\tablefootnote{\url{http://www.bfs.admin.ch/}} \\
		\emph{FAO} & \emph{Food and Agriculture Organization of the United Nations}\tablefootnote{\url{http://www.fao.org/home/en/}} \\
		\emph{WB} & \emph{World Bank}\tablefootnote{\url{http://www.worldbank.org/}} \\
		\emph{FRB} & \emph{Federal Reserve Board}\tablefootnote{\url{http://www.federalreserve.gov/}} \\
		\emph{TI} & \emph{Transparency International}\tablefootnote{\url{http://www.transparency.org/}} \\
		\emph{OECD} & \emph{Organisation for Economic Co-operation and Development}\tablefootnote{\url{http://www.oecd.org/}} \\
		\emph{BIS} & \emph{Bank for International Settlements}\tablefootnote{\url{http://www.bis.org/}} \\
		\emph{ABS} & \emph{Australian Bureau of Statistics}\tablefootnote{\url{http://abs.gov.au/}} \\
		\emph{IEEE-VIS} & \emph{IEEE VIS Source Data} \\
		\emph{ACORN-SAT} & \emph{Australian Climate Observations Reference Network - Surface Air Temperature Dataset} \\
		\emph{HDP} & \emph{HealthData.gov Platform (HDP) on the Semantic Web} \\
		\emph{Eurostat} & \emph{The Eurostat Linked Data} (SPARQL endpoint unavailable) \\
		\emph{Asturias} & \emph{Nomenclator Asturias} (SPARQL endpoint unavailable!) \\
		\emph{ISTAT} & \emph{ISTAT Immigration (LinkedOpenData.it)} (SPARQL endpoint unavailable) \\
		\emph{ICANE} & \emph{Statistical Office of Cantabria (Instituto Cántabro de Estadística, ICANE)} \\ 
		      & (SPARQL endpoint unavailable) \\
		\emph{EE-2009} & \emph{European Election Results 2009} (SPARQL endpoint unavailable) \\
		\emph{EU-B} & \emph{Standard Eurobarometer} (SPARQL endpoint unavailable) \\
		\emph{ECB-S} & \emph{European Central Bank Statistics (PublicData.eu)} (SPARQL endpoint unavailable) \\
		\emph{CPV-2008} & \emph{Common Procurement Vocabulary (CPV) 2008} (SPARQL endpoint unavailable) \\
		\emph{CPV-2003} & \emph{Common Procurement Vocabulary (CPV) 2003} (SPARQL endpoint unavailable) \\
		\end{tabular}
	\caption{QB Data Sets Abbreviations}
	\label{tab:data-cube-data-sets-abbreviations}
\end{table}

\begin{table}[H]
    \begin{center}
    \begin{tabular}{@{}lccccccccccc@{}}
           & \multicolumn{5}{c}{\textbf{Counts}}
    \\  \cmidrule{2-6}
    \\       \textbf{Data Sets}
           & \textbf{\rot{triples}}
           & \textbf{\rot{qb:DataSet}}
           & \textbf{\rot{qb:DataStructureDefinition}}
           & \textbf{\rot{qb:Observation}}
           & \textbf{\rot{qb:Slice}}
    \\ \midrule
    \emph{ECB} & 468,899,474 & 55 & 46 & $>$11,000,000 & 428,698 \\
		\emph{UIS} & 10,400,534 & 5 & 5 & 1,437,651 & 0 &  \\
		\emph{IMF} & 35,688,446 & 4 & 8 & 3,603,719 & 0 &  \\
		\emph{BFS} & 1,533,743 & 0 & 0 & 8 & 0 &  \\
		\emph{FAO} & 53,000,000 & 10 & 10 & $>$7,100,000 & 0 \\
		\emph{WB} & 174,006,552 & 9,466 & 59 & $>$17,000,000 & 0 \\
		\emph{FRB} & 185,266,900 & 49 & 98 & $>$9,500,000 & 0 \\
		\emph{TI} & 52,233 & 6 & 6 & 3,928 & 0 \\
		\emph{OECD} & 304,995,160 & 136 & 140 & $>$12,000,000 & 0 \\
		\emph{BIS} & 54,197,482 & 6 & 12 & 3,606,466 & 47,914 \\
		\emph{ABS} & 2,357,400,000 & 253 & 257 & $>$11,000,000 & 0  \\
		\emph{IEEE-VIS} & 19,935,340 & 0 & 0 & 1,350 & 0 \\
		\emph{ACORN-SAT} & 98,381,319 & 0 & 4 & 0 & 0 \\
		\emph{HDP} & 12,226,427 & 0 & 0 & 0 & 0 \\
		\hline
		\textbf{Total} & 3,775,983,610 & 9,990 & & & & \\
    \bottomrule
    \end{tabular}
    \caption{QB Data Sets Overview}
		\label{tab:data-cube-overview}
    \end{center}
\end{table}

\begin{table}[H]
	\centering
		\begin{tabular}{l|l}
      \textbf{Data Sets} & \textbf{SPARQL Endpoints} \\		
      \hline
      \emph{ECB} & \url{http://ecb.270a.info/sparql} \\
			\emph{UIS} & \url{http://uis.270a.info/sparql} \\
			\emph{IMF} & \url{http://imf.270a.info/sparql} \\
			\emph{BFS} & \url{http://bfs.270a.info/sparql} \\
			\emph{FAO} & \url{http://fao.270a.info/sparql} \\
			\emph{WB} & \url{http://worldbank.270a.info/sparql} \\
			\emph{FRB} & \url{http://frb.270a.info/sparql} \\
			\emph{TI} & \url{http://transparency.270a.info/sparql} \\
			\emph{OECD} & \url{http://oecd.270a.info/sparql} \\
			\emph{BIS} & \url{http://bis.270a.info/sparql} \\
			\emph{ABS} & \url{http://abs.270a.info/sparql} \\
			\emph{ACORN-SAT} & \url{http://lab.environment.data.gov.au/sparql} \\
			\emph{HDP} & \url{http://healthdata.tw.rpi.edu/sparql} \\
		\end{tabular}
	\caption{QB SPARQL Endpoints}
	\label{tab:data-cube-sparql-endpoints}
\end{table}

\subsection{Detailed Evaluation}

In this sub-section, we give details about the evaluation in form of diverse tables containing the number of constraint violations per evaluated data set and constraint of particular constraint types.

\begin{table}[H]
    \begin{center}
    \begin{tabular}{@{}lccccccc@{}}
           & \multicolumn{7}{c}{\textbf{Data Sets}}
    \\  \cmidrule{2-8}
    \\       \textbf{Data Model Consistency}
           & \rot{\emph{ECB}}
           & \rot{\emph{UIS}}
           & \rot{\emph{IMF}}
           & \rot{\emph{BFS}}
           & \rot{\emph{FAO}}
					 & \rot{\emph{WB}}
					 & \rot{\emph{FRB}}
    \\ \midrule
    \emph{DATA-MODEL-CONSISTENCY-01}\textsuperscript{**} & $\checkmark$ (2) & $\checkmark$ & $\checkmark$ & $\checkmark$ & $\checkmark$ & $\checkmark$ & $\checkmark$ \\
    \emph{DATA-MODEL-CONSISTENCY-02}\textsuperscript{***} & $\checkmark$ (2) & $\checkmark$ & $\checkmark$ & $\checkmark$ & $\checkmark$ & $\checkmark$ & $\checkmark$ \\
    \emph{DATA-MODEL-CONSISTENCY-03}\textsuperscript{***} & $\checkmark$ (2) & $\checkmark$ & $\checkmark$ & $\checkmark$ & $\checkmark$ & $\checkmark$ & $\checkmark$ \\
    \emph{DATA-MODEL-CONSISTENCY-04}\textsuperscript{***} & $\checkmark$ (6) & $\checkmark$ & $\checkmark$ & $\checkmark$ & $\checkmark$ & $\checkmark$ & 14,372 \\
		\emph{DATA-MODEL-CONSISTENCY-05}\textsuperscript{**} & 1,198,352 (50) & \ding{55} & \ding{55} & $\checkmark$ & \ding{55} & $\checkmark$ & 16,175,814 (42) \\
		\emph{DATA-MODEL-CONSISTENCY-06}\textsuperscript{***} & $\checkmark$ (2) & $\checkmark$ & $\checkmark$ & $\checkmark$ & $\checkmark$ & $\checkmark$ & $\checkmark$ \\
		\emph{DATA-MODEL-CONSISTENCY-07}\textsuperscript{***} & $\checkmark$ (9) & $\checkmark$ & 99,091 & $\checkmark$ & $\checkmark$ & $\checkmark$ & $\checkmark$ (1) \\
		\emph{DATA-MODEL-CONSISTENCY-08}\textsuperscript{***} & $\checkmark$ (2) & $\checkmark$ & $\checkmark$ & $\checkmark$ & $\checkmark$ & $\checkmark$ & $\checkmark$ \\
		\emph{DATA-MODEL-CONSISTENCY-09}\textsuperscript{***} & $\checkmark$ (2) & $\checkmark$ & $\checkmark$ & $\checkmark$ & $\checkmark$ & $\checkmark$ & $\checkmark$ \\
		\emph{DATA-MODEL-CONSISTENCY-10}\textsuperscript{***} (!) & - & - & - & - & - & - & - \\
		\emph{DATA-MODEL-CONSISTENCY-11}\textsuperscript{**} & 6,511 (10) & $\checkmark$ & $\checkmark$ & $\checkmark$ & $\checkmark$ & $\checkmark$ & $\checkmark$ \\
    \bottomrule
    \end{tabular}
    \caption{Evaluation of QB Data Sets - Data Model Consistency (1)}
		\label{evaluation-of-data-cube-data-sets-data-model-consistency-1}
    \end{center}
\end{table}

\begin{table}[H]
    \begin{center}
    \begin{tabular}{@{}lccccccc@{}}
           & \multicolumn{7}{c}{\textbf{Data Sets}}
    \\  \cmidrule{2-8}
    \\       \textbf{Data Model Consistency}
					 & \rot{\emph{TI}}
					 & \rot{\emph{OECD}}
					 & \rot{\emph{BIS}}
					 & \rot{\emph{ABS}}
					 & \rot{\emph{IEEE-VIS}}
					 & \rot{\emph{ACORN-SAT}}
					 & \rot{\emph{HDP}}
    \\ \midrule
    \emph{DATA-MODEL-CONSISTENCY-01}\textsuperscript{**} & $\checkmark$ & $\checkmark$ & $\checkmark$ & $\checkmark$ & $\checkmark$ & $\checkmark$ & $\checkmark$ \\
    \emph{DATA-MODEL-CONSISTENCY-02}\textsuperscript{***} & $\checkmark$ & $\checkmark$ & $\checkmark$ & $\checkmark$ & $\checkmark$ & 8 & $\checkmark$ \\
    \emph{DATA-MODEL-CONSISTENCY-03}\textsuperscript{***} & $\checkmark$ & $\checkmark$ & $\checkmark$ & $\checkmark$ & $\checkmark$ & $\checkmark$ & $\checkmark$ \\
    \emph{DATA-MODEL-CONSISTENCY-04}\textsuperscript{***} & $\checkmark$ & $\checkmark$& $\checkmark$ & $\checkmark$ (6) & $\checkmark$ & $\checkmark$ & $\checkmark$ \\
		\emph{DATA-MODEL-CONSISTENCY-05}\textsuperscript{**} & $\checkmark$ & 21,142,838 (116) & \ding{55} &  6,997,098 (246) & $\checkmark$ & $\checkmark$ & $\checkmark$ \\
		\emph{DATA-MODEL-CONSISTENCY-06}\textsuperscript{***} & $\checkmark$ & $\checkmark$ & $\checkmark$ & $\checkmark$ & $\checkmark$ & $\checkmark$ & $\checkmark$ \\
		\emph{DATA-MODEL-CONSISTENCY-07}\textsuperscript{***} & $\checkmark$ & $\checkmark$ & $\checkmark$ & $\checkmark$ (8) & $\checkmark$ & $\checkmark$ & $\checkmark$ \\
		\emph{DATA-MODEL-CONSISTENCY-08}\textsuperscript{***} & $\checkmark$ & $\checkmark$ & $\checkmark$ & $\checkmark$ & $\checkmark$ & $\checkmark$ & $\checkmark$ \\
		\emph{DATA-MODEL-CONSISTENCY-09}\textsuperscript{***} & $\checkmark$ & $\checkmark$ & $\checkmark$ & $\checkmark$ & $\checkmark$ & $\checkmark$ & $\checkmark$ \\
		\emph{DATA-MODEL-CONSISTENCY-10}\textsuperscript{***} (!) & - & - & - & - & - & - & - \\
		\emph{DATA-MODEL-CONSISTENCY-11}\textsuperscript{**} & $\checkmark$ & $\checkmark$ & $\checkmark$ & $\checkmark$ & $\checkmark$ & $\checkmark$ & $\checkmark$ \\
    \bottomrule
    \end{tabular}
    \caption{Evaluation of QB Data Sets - Data Model Consistency (2)}
		\label{evaluation-of-data-cube-data-sets-data-model-consistency-1}
    \end{center}
\end{table}

\begin{table}[H]
    \begin{center}
    \begin{tabular}{@{}lcccccccccccccc@{}}
           & \multicolumn{14}{c}{\textbf{Data Sets}}
    \\  \cmidrule{2-15}
    \\       \textbf{Existential Quantifications}
           & \rot{\emph{ECB}}
           & \rot{\emph{UIS}}
           & \rot{\emph{IMF}}
           & \rot{\emph{BFS}}
           & \rot{\emph{FAO}}
					 & \rot{\emph{WB}}
					 & \rot{\emph{FRB}}
					 & \rot{\emph{TI}}
					 & \rot{\emph{OECD}}
					 & \rot{\emph{BIS}}
					 & \rot{\emph{ABS}}
					 & \rot{\emph{IEEE-VIS}}
					 & \rot{\emph{ACORN-SAT}}
					 & \rot{\emph{HDP}}
    \\ \midrule
		\emph{EXISTENTIAL-QUANTIFICATIONS-01}\textsuperscript{***} & 9 & $\checkmark$ & 11 & 7 & 8 & 77 & 8 & 9 & 7 & 8 & 7 & $\checkmark$ & $\checkmark$ & $\checkmark$ \\
		\emph{EXISTENTIAL-QUANTIFICATIONS-02}\textsuperscript{***} & $\checkmark$ & $\checkmark$ & $\checkmark$ & $\checkmark$ & $\checkmark$ & $\checkmark$ & $\checkmark$ & $\checkmark$ & $\checkmark$ & $\checkmark$ & $\checkmark$ & $\checkmark$ & $\checkmark$ & $\checkmark$ \\
		\emph{EXISTENTIAL-QUANTIFICATIONS-03}\textsuperscript{***} & $\checkmark$ & $\checkmark$ & $\checkmark$ & $\checkmark$ & $\checkmark$ & 59 & $\checkmark$ & 6 & $\checkmark$ & $\checkmark$ & $\checkmark$ & $\checkmark$ & 4 & $\checkmark$ \\
		\emph{EXISTENTIAL-QUANTIFICATIONS-04}\textsuperscript{***} & $\checkmark$ & $\checkmark$ & $\checkmark$ & $\checkmark$ & $\checkmark$ & $\checkmark$ & $\checkmark$ & $\checkmark$ & $\checkmark$ & $\checkmark$ & $\checkmark$ & $\checkmark$ & $\checkmark$ & $\checkmark$ \\
    \bottomrule
    \end{tabular}
    \caption{Evaluation of QB Data Sets - Existential Quantifications}
    \end{center}
\end{table}

\begin{table}[H]
    \begin{center}
    \begin{tabular}{@{}lcccccccccc@{}}
           & \multicolumn{10}{c}{\textbf{Data Sets}}
    \\  \cmidrule{2-11}
    \\       \textbf{Cardinality Restrictions}
           & \rot{\emph{ECB}}
           & \rot{\emph{UIS}}
           & \rot{\emph{IMF}}
           & \rot{\emph{BFS}}
           & \rot{\emph{FAO}}
					 & \rot{\emph{WB}}
					 & \rot{\emph{FRB}}
					 & \rot{\emph{TI}}
					 & \rot{\emph{OECD}}
					 & \rot{\emph{BIS}}
    \\ \midrule
		\emph{MINIMUM-QUALIFIED-CARDINALITY-RESTRICTIONS-01 (!)}\textsuperscript{***} & - & - & - & - & - & - & - & - & - & - \\
		\emph{MINIMUM-QUALIFIED-CARDINALITY-RESTRICTIONS-02}\textsuperscript{***} & \ding{55} & 118 & 8 & 8 & 30 & $\checkmark$ & 30 & $\checkmark$ & \ding{55} & 12 \\
		\emph{MAXIMUM-QUALIFIED-CARDINALITY-RESTRICTIONS-01}\textsuperscript{***} & $\checkmark$ & $\checkmark$ & $\checkmark$ & $\checkmark$ & $\checkmark$ & $\checkmark$ & $\checkmark$ & $\checkmark$ & $\checkmark$ & $\checkmark$ \\
		\emph{EXACT-UNQUALIFIED-CARDINALITY-RESTRICTIONS-01}\textsuperscript{***} & $\checkmark$ & $\checkmark$ & $\checkmark$ & $\checkmark$ & $\checkmark$ & $\checkmark$ & $\checkmark$ & $\checkmark$ & $\checkmark$ & $\checkmark$ \\
		\emph{EXACT-QUALIFIED-CARDINALITY-RESTRICTIONS-02}\textsuperscript{***} & $\checkmark$ & $\checkmark$ & $\checkmark$ & $\checkmark$ & $\checkmark$ & 1 & $\checkmark$ & $\checkmark$ & $\checkmark$ & $\checkmark$ \\
    \bottomrule
    \end{tabular}
    \caption{Evaluation of QB Data Sets - Cardinality Restrictions (1)}
    \end{center}
\end{table}

\begin{table}[H]
    \begin{center}
    \begin{tabular}{@{}lcccc@{}}
           & \multicolumn{4}{c}{\textbf{Data Sets}}
    \\  \cmidrule{2-5}
    \\       \textbf{Cardinality Restrictions}
					 & \rot{\emph{ABS}}
					 & \rot{\emph{IEEE-VIS}}
					 & \rot{\emph{ACORN-SAT}}
					 & \rot{\emph{HDP}}
    \\ \midrule
		\emph{MINIMUM-QUALIFIED-CARDINALITY-RESTRICTIONS-01 (!)}\textsuperscript{***} & - & - & - & - \\
		\emph{MINIMUM-QUALIFIED-CARDINALITY-RESTRICTIONS-02}\textsuperscript{***} & \ding{55} & 1,350 & $\checkmark$ & $\checkmark$ \\
		\emph{MAXIMUM-QUALIFIED-CARDINALITY-RESTRICTIONS-01}\textsuperscript{***} & $\checkmark$ (2) & $\checkmark$ & $\checkmark$ & $\checkmark$ \\
		\emph{EXACT-UNQUALIFIED-CARDINALITY-RESTRICTIONS-01}\textsuperscript{***} & $\checkmark$ & $\checkmark$ & $\checkmark$ & $\checkmark$ \\
		\emph{EXACT-QUALIFIED-CARDINALITY-RESTRICTIONS-02}\textsuperscript{***} & $\checkmark$ & $\checkmark$ & $\checkmark$ & $\checkmark$ \\
    \bottomrule
    \end{tabular}
    \caption{Evaluation of QB Data Sets - Cardinality Restrictions (2)}
    \end{center}
\end{table}

\begin{table}[H]
    \begin{center}
    \begin{tabular}{@{}lcccccccccccccc@{}}
           & \multicolumn{14}{c}{\textbf{Data Sets}}
    \\  \cmidrule{2-15}
    \\       \textbf{Structure}
           & \rot{\emph{ECB}}
           & \rot{\emph{UIS}}
           & \rot{\emph{IMF}}
           & \rot{\emph{BFS}}
           & \rot{\emph{FAO}}
					 & \rot{\emph{WB}}
					 & \rot{\emph{FRB}}
					 & \rot{\emph{TI}}
					 & \rot{\emph{OECD}}
					 & \rot{\emph{BIS}}
					 & \rot{\emph{ABS}}
					 & \rot{\emph{IEEE-VIS}}
					 & \rot{\emph{ACORN-SAT}}
					 & \rot{\emph{HDP}}
    \\ \midrule
		\emph{STRUCTURE-01}\textsuperscript{***} & $\checkmark$ & $\checkmark$ & $\checkmark$ & $\checkmark$ & $\checkmark$ & $\checkmark$ & $\checkmark$ & $\checkmark$ & $\checkmark$ & $\checkmark$ & $\checkmark$ & $\checkmark$ & $\checkmark$ & $\checkmark$ \\
		\emph{STRUCTURE-02}\textsuperscript{***} & $\checkmark$ & $\checkmark$ & $\checkmark$ & $\checkmark$ & $\checkmark$ & $\checkmark$ & $\checkmark$ & $\checkmark$ & $\checkmark$ & $\checkmark$ & $\checkmark$ & $\checkmark$ & $\checkmark$ & $\checkmark$ \\
    \bottomrule
    \end{tabular}
    \caption{Evaluation of QB Data Sets - Structure}
    \end{center}
\end{table}

\begin{table}[H]
    \begin{center}
    \begin{tabular}{@{}lcccccccccccccc@{}}
           & \multicolumn{14}{c}{\textbf{Data Sets}}
    \\  \cmidrule{2-15}
    \\       \textbf{Constraints}
           & \rot{\emph{ECB}}
           & \rot{\emph{UIS}}
           & \rot{\emph{IMF}}
           & \rot{\emph{BFS}}
           & \rot{\emph{FAO}}
					 & \rot{\emph{WB}}
					 & \rot{\emph{FRB}}
					 & \rot{\emph{TI}}
					 & \rot{\emph{OECD}}
					 & \rot{\emph{BIS}}
					 & \rot{\emph{ABS}}
					 & \rot{\emph{IEEE-VIS}}
					 & \rot{\emph{ACORN-SAT}}
					 & \rot{\emph{HDP}}
    \\ \midrule
		\emph{PROPERTY-DOMAIN-01 (!)}\textsuperscript{***} &  &  &  &  &  &  &  &  &  &  &  &  \\
		\emph{PROPERTY-RANGES-01 (!)}\textsuperscript{***} &  &  &  &  &  &  &  &  &  &  &  &  \\
		\emph{DISJOINT-PROPERTIES-01 (!)}\textsuperscript{***} &  &  &  &  &  &  &  &  &  &  &  &  \\
		\emph{DISJOINT-CLASSES-01 (!)}\textsuperscript{***} &  &  &  &  &  &  &  &  &  &  &  &  \\
		\emph{EQUIVALENT-PROPERTIES-01 (!)}\textsuperscript{*} &  &  &  &  &  &  &  &  &  &  &  &  \\
		\emph{UNIVERSAL-QUANTIFICATIONS-01 (!)}\textsuperscript{***} &  &  &  &  &  &  &  &  &  &  &  &  \\
		\emph{MEMBERSHIP-IN-CONTROLLED-VOCABULARIES-01 (!)}\textsuperscript{***} &  &  &  &  &  &  &  &  &  &  &  &  \\
		\emph{CONTEXT-SPECIFIC-VALID-CLASSES-01 (!)}\textsuperscript{*} &  &  &  &  &  &  &  &  &  &  &  &  \\
		\emph{CONTEXT-SPECIFIC-VALID-PROPERTIES-01 (!)}\textsuperscript{*} &  &  &  &  &  &  &  &  &  &  &  &  \\
		\emph{RECOMMENDED-PROPERTIES-01 (!)}\textsuperscript{*} &  &  &  &  &  &  &  &  &  &  &  &  \\
		\emph{VALUE-IS-VALID-FOR-DATATYPE-01 (!)}\textsuperscript{***} &  &  &  &  &  &  &  &  &  &  &  &  \\
		\emph{VOCABULARY-01 (!)}\textsuperscript{***} &  &  &  &  &  &  &  &  &  &  &  &  \\
    \bottomrule
    \end{tabular}
    \caption{Evaluation of QB Data Sets - Constraints (1)}
    \end{center}
\end{table}

\begin{table}[H]
    \begin{center}
    \begin{tabular}{@{}lcccccccccccccc@{}}
           & \multicolumn{14}{c}{\textbf{Data Sets}}
    \\  \cmidrule{2-15}
    \\       \textbf{Constraints}
           & \rot{\emph{ECB}}
           & \rot{\emph{UIS}}
           & \rot{\emph{IMF}}
           & \rot{\emph{BFS}}
           & \rot{\emph{FAO}}
					 & \rot{\emph{WB}}
					 & \rot{\emph{FRB}}
					 & \rot{\emph{TI}}
					 & \rot{\emph{OECD}}
					 & \rot{\emph{BIS}}
					 & \rot{\emph{ABS}}
					 & \rot{\emph{IEEE-VIS}}
					 & \rot{\emph{ACORN-SAT}}
					 & \rot{\emph{HDP}}
    \\ \midrule
		\emph{HTTP-URI-SCHEME-VIOLATION (!)}\textsuperscript{***} \\
    \bottomrule
    \end{tabular}
    \caption{Evaluation of QB Data Sets - Constraints (2)}
    \end{center}
\end{table}

\section{Evaluation of Metadata on Thesauri (SKOS)}

In this section, the quality of the metadata on thesauri (SKOS) is evaluated by validating appropriate RDF constraints assigned to several RDF constraint types.
First, we give an overview on the evaluated thesauri and finally we provide details about the evaluation.

\subsection{Data Sets Overview}

There is a website giving an overview on available SKOS data sets\footnote{\url{http://datahub.io/de/dataset?tags=format-skos}}
and another one giving an overview on available thesauri\footnote{\url{http://datahub.io/de/dataset?tags=thesaurus}}.
Tables \ref{tab:thesauri-abbreviations} and \ref{tab:thesauri-sparql-endpoints} give an overview on the evaluated thesauri, their abbreviations, and publicly available SPARQL endpoints.
Table \ref{tab:thesauri-overview} comprehends the number of triples, data sets, and instances of multiple vocabulary-specific classes.

\begin{table}[H]
	\centering
		\begin{tabular}{l|l}
      \textbf{Abbr.} & \textbf{Thesauri} \\		
      \hline
    \emph{TheSoz} & \emph{Thesaurus for the Social Sciences}\tablefootnote{\url{http://www.ecb.europa.eu/home/html/index.en.html}} \\
	  \emph{STW} & \emph{Thesaurus for Economics}\tablefootnote{\url{http://zbw.eu/stw/versions/latest/about}} \\
	  \emph{AGROVOC} & \emph{AGROVOC Multilingual agricultural thesaurus}\tablefootnote{\url{http://202.45.139.84:10035/catalogs/fao/repositories/agrovoc}} \\
		\emph{UNESCO} & \emph{UNESCO Thesaurus}\tablefootnote{\url{http://skos.um.es/sparql/}} \\
		\emph{TGN} & \emph{The Getty Thesaurus of Geographic Names}\tablefootnote{\url{http://vocab.getty.edu/sparql}} \\
		\emph{EARTh} & \emph{Environmental Applications Reference Thesaurus}\tablefootnote{\url{http://linkeddata.ge.imati.cnr.it/resource/EARTh/}} \\
		\emph{ODT} & \emph{Open Data Thesaurus}\tablefootnote{\url{http://vocabulary.semantic-web.at/PoolParty/wiki/OpenData}} \\
		\emph{SLD} & \emph{Spanish Linguistic Datasets}\tablefootnote{\url{http://linguistic.linkeddata.es}} \\
		\emph{SSWT} & \emph{Social Semantic Web Thesaurus}\tablefootnote{\url{http://vocabulary.semantic-web.at/PoolParty/wiki/semweb}} \\
		\emph{GBA-GU} & \emph{Thesaurus of the Geological Survey of Austria (GBA) - Geology Unit}\tablefootnote{\url{http://resource.geolba.ac.at/}} \\
		\emph{GBA-GTS} & \emph{Thesaurus of the Geological Survey of Austria (GBA) - Geologic Time Scale}\tablefootnote{\url{http://resource.geolba.ac.at/}} \\
		\emph{GBA-L} & \emph{Thesaurus of the Geological Survey of Austria (GBA) - Lithology}\tablefootnote{\url{http://resource.geolba.ac.at/}} \\
		\emph{GBA-LU} & \emph{Thesaurus of the Geological Survey of Austria (GBA) - Lithotectonic Unit}\tablefootnote{\url{http://resource.geolba.ac.at/}} \\
		\emph{GEMET} & \emph{GEneral Multilingual Environmental Thesaurus}\tablefootnote{\url{http://www.eionet.europa.eu/gemet/}} \\
		\emph{EuroVoc} & \emph{EuroVoc}\tablefootnote{\url{http://open-data.europa.eu/de/data/dataset/eurovoc}} \\
		\emph{CECCT} & \emph{Clean Energy and Climate Change Thesaurus}\tablefootnote{\url{http://data.reegle.info/thesaurus/guide}} \\
		\end{tabular}
	\caption{Thesauri Abbreviations}
	\label{tab:thesauri-abbreviations}
\end{table}

\begin{table}[H]
    \begin{center}
    \begin{tabular}{@{}lccccccccccc@{}}
           & \multicolumn{7}{c}{\textbf{Counts}}
    \\  \cmidrule{2-8}
    \\       \textbf{Thesauri}
           & \textbf{\rot{triples}}
           & \textbf{\rot{skos:ConceptScheme}}
           & \textbf{\rot{sko:Concept}}
           & \textbf{\rot{skos:broader}}
           & \textbf{\rot{skos:narrower}}
					 & \textbf{\rot{skos:hasTopConcept}}
					 & \textbf{\rot{skos:inScheme}}
    \\ \midrule
    \emph{TheSoz} & 439,153 & 1 & 8,426 & 13,705 & 13,706 & 0 & 48,529 \\
	  \emph{STW} & 221,668 & 1 & 13,468 & 13,732 & 13732 & 7 & 13,180 \\
	  \emph{AGROVOC} & 6,080,477 & 1 & 32,310 & 33,507 & 33,507 & 25 & 32,310 \\
		\emph{UNESCO} & 288,346 & 9 & 26,714 & 20,028 & 20,028 & 607 & 32,009 \\
		\emph{TGN} & 16,112,321 & 8 & 2,898,775 & 0 & 0 & 0 & 1,453,767\\
		\emph{EARTh} & 9,287,364 & 11 & 295,375 & 288,208 & 93,827 & 479 & 295,376 \\
		\emph{ODT} & 3,290 & 6 & 108 & 93 & 93 & 30 & 0 \\
		\emph{SLD} & 7,629,211 & 0 & 31,195 & 0 & 0 & 0 & 0 \\
		\emph{SSWT} & 64,698 & 9 & 2,127 & 2,300 & 2,301 & 38 & 0 \\
		\emph{GBA-GU} & 25,718 & 3 & 878 & 1,005 & 1,005 & 14 & 0 \\
		\emph{GBA-GTS} & 7,875 & 3 & 213 & 208 & 208 & 5 & 0 \\
		\emph{GBA-L} & 9,317 & 1 & 249 & 249 & 249 & 4 & 0 \\
		\emph{GBA-LU} & 9,504 & 3 & 364 & 359 & 359 & 7 & 0 \\
		\emph{GEMET} & 372,889,229 & 3,680 & 414,659 & 62,193 & 21,685 & 30,806 & 409,290 \\
		\emph{EuroVoc} & 64,477,774 & 439 & 79,557 & 6,922 & 0 & 532 & 14,428 \\
		\emph{CECCT} & 191,336 & 3 & 3,419 & 3,761 & 3,762 & 28 & 0 \\
		\hline
		\textbf{Total} & 477,737,281 & 4,178 & & & & \\
    \bottomrule
    \end{tabular}
    \caption{Thesauri Overview}
		\label{tab:thesauri-overview}
    \end{center}
\end{table}

\begin{table}[H]
	\centering
		\begin{tabular}{l|l}
      \textbf{Thesauri} & \textbf{SPARQL Endpoints} \\		
      \hline
      \emph{TheSoz} & \url{http://lod.gesis.org/thesoz/sparql} \\
			\emph{STW} & \url{http://zbw.eu/beta/sparql/stw/query} \\
			\emph{AGROVOC} & \url{http://202.45.139.84:10035/catalogs/fao/repositories/agrovoc} \\
			\emph{UNESCO} & \url{http://skos.um.es/sparql/} \\
			\emph{TGN} & \url{http://vocab.getty.edu/} \\
			\emph{EARTh} & \url{http://linkeddata.ge.imati.cnr.it:8890/sparql} \\
			\emph{ODT} & \url{http://vocabulary.semantic-web.at/PoolParty/sparql/OpenData} \\
			\emph{SLD} & \url{http://linguistic.linkeddata.es/sparql} \\
			\emph{SSWT} & \url{http://vocabulary.semantic-web.at/PoolParty/sparql/semweb} \\
			\emph{GBA-GU} & \url{http://resource.geolba.ac.at/PoolParty/sparql/GeologicUnit} \\
			\emph{GBA-GTS} & \url{http://resource.geolba.ac.at/PoolParty/sparql/GeologicTimeScale} \\
			\emph{GBA-L} & \url{http://resource.geolba.ac.at/PoolParty/sparql/lithology} \\
			\emph{GBA-LU} & \url{http://resource.geolba.ac.at/PoolParty/sparql/tectonicunit} \\
			\emph{GEMET} & \url{http://semantic.eea.europa.eu/sparql} \\
			\emph{EuroVoc} & \url{http://open-data.europa.eu/de/linked-data} \\
			\emph{CECCT} & \url{http://poolparty.reegle.info/PoolParty/sparql/glossary}
		\end{tabular}
	\caption{Thesauri SPARQL Endpoints}
	\label{tab:thesauri-sparql-endpoints}
\end{table}

\subsection{Detailed Evaluation}

In this sub-section, we give details about the evaluation in form of diverse tables containing the number of constraint violations per evaluated data set and constraint of particular constraint types.

\begin{table}[H]
    \begin{center}
    \begin{tabular}{@{}lcccccccccccc@{}}
           & \multicolumn{12}{c}{\textbf{Data Sets}}
    \\  \cmidrule{2-13}
    \\       \textbf{Data Model Consistency}
           & \emph{\rot{TheSoz}}
           & \emph{\rot{STW}}
           & \emph{\rot{AGROVOC}}
					 & \emph{\rot{TGN}}
           & \emph{\rot{UNESCO}}
					 & \emph{\rot{ODT}}
					 & \emph{\rot{SSWT}}
					 & \emph{\rot{GBA-GU}}
					 & \emph{\rot{GBA-GTS}}
					 & \emph{\rot{GBA-L}}
					 & \emph{\rot{GBA-LU}}
					 & \emph{\rot{CECCT}}
    \\ \midrule
    \emph{DATA-MODEL-CONSISTENCY-01 (!)}\textsuperscript{*} &  &  &  &  &  & \\
    \emph{DATA-MODEL-CONSISTENCY-02 (!)}\textsuperscript{*} &  &  &  &  &  & \\
    \emph{DATA-MODEL-CONSISTENCY-03 (!)}\textsuperscript{*} &  &  &  &  &  & \\
    \bottomrule
    \end{tabular}
    \caption{Thesauri Evaluation - Data Model Consistency (1)}
		\label{tab:thesauri-evaluation-data-model-consistency-1}
    \end{center}
\end{table}

\begin{table}[H]
    \begin{center}
    \begin{tabular}{@{}lcccc@{}}
           & \multicolumn{4}{c}{\textbf{Data Sets}}
    \\  \cmidrule{2-5}
    \\       \textbf{Data Model Consistency}
					 & \emph{\rot{EARTh}}
					 & \emph{\rot{GEMET}}
					 & \emph{\rot{EuroVoc}}
					 & \emph{\rot{SLD}}
    \\ \midrule
    \emph{DATA-MODEL-CONSISTENCY-01 (!)}\textsuperscript{*} \\
    \emph{DATA-MODEL-CONSISTENCY-02 (!)}\textsuperscript{*} \\
    \emph{DATA-MODEL-CONSISTENCY-03 (!)}\textsuperscript{*} \\
    \bottomrule
    \end{tabular}
    \caption{Thesauri Evaluation - Data Model Consistency (2)}
		\label{tab:thesauri-evaluation-data-model-consistency-2}
    \end{center}
\end{table}

\begin{table}[H]
    \begin{center}
    \begin{tabular}{@{}lcccccccccccc@{}}
           & \multicolumn{12}{c}{\textbf{Data Sets}}
    \\  \cmidrule{2-13}
    \\       \textbf{Labeling and Documentation}
           & \emph{\rot{TheSoz}}
           & \emph{\rot{STW}}
           & \emph{\rot{AGROVOC}}
					 & \emph{\rot{TGN}}
           & \emph{\rot{UNESCO}}
					 & \emph{\rot{ODT}}
					 & \emph{\rot{SSWT}}
					 & \emph{\rot{GBA-GU}}
					 & \emph{\rot{GBA-GTS}}
					 & \emph{\rot{GBA-L}}
					 & \emph{\rot{GBA-LU}}
					 & \emph{\rot{CECCT}}
    \\ \midrule
		\emph{LABELING-AND-DOCUMENTATION-01}\textsuperscript{*} & 8,426 & 11,508 & 19,829 & 1,110 & \ding{55} & 36 & 1,475 & 5 & 2 & $\checkmark$ & 107 & 486 \\
		\emph{LABELING-AND-DOCUMENTATION-02}\textsuperscript{*} & $>$1 & \ding{55} & $>$100 & 287 & \ding{55} & $\checkmark$ & $\checkmark$ & $\checkmark$ & $\checkmark$ & $\checkmark$ & $\checkmark$ & $\checkmark$ \\
		\emph{LABELING-AND-DOCUMENTATION-03}\textsuperscript{*} & $\checkmark$ & $\checkmark$ & 1 & 14,114 & \ding{55} & $\checkmark$ & $\checkmark$ & 1 & $\checkmark$ & $\checkmark$ & 1 & $\checkmark$ \\
		\emph{LABELING-AND-DOCUMENTATION-04 (!)}\textsuperscript{*} &  &  &  &  &  & & &  \\
		\emph{LABELING-AND-DOCUMENTATION-05}\textsuperscript{*} & $\checkmark$ & $\checkmark$ & 4 & $\checkmark$ & 1 & 2 & 2 & 1 & $\checkmark$ & $\checkmark$ & $\checkmark$ & 7 \\
		\emph{LABELING-AND-DOCUMENTATION-06}\textsuperscript{*} & 975,340 & $\checkmark$ & $\checkmark$ & 2 & $\checkmark$ & $\checkmark$ & $\checkmark$ & $\checkmark$ & $\checkmark$ & $\checkmark$ & $\checkmark$ & $\checkmark$\\
    \bottomrule
    \end{tabular}
    \caption{Thesauri Evaluation - Labeling and Documentation (1)}
		\label{tab:thesauri-evaluation-labeling-and-documentation-1}
    \end{center}
\end{table}

\begin{table}[H]
    \begin{center}
    \begin{tabular}{@{}lcccc@{}}
           & \multicolumn{4}{c}{\textbf{Data Sets}}
    \\  \cmidrule{2-5}
    \\       \textbf{Labeling and Documentation}
					 & \emph{\rot{EARTh}}
					 & \emph{\rot{GEMET}}
					 & \emph{\rot{EuroVoc}}
					 & \emph{\rot{SLD}}
    \\ \midrule
		\emph{LABELING-AND-DOCUMENTATION-01}\textsuperscript{*}& 264,687 & \ding{55} & 54,911 & 31,195 \\
		\emph{LABELING-AND-DOCUMENTATION-02}\textsuperscript{*} & \ding{55} & \ding{55} & \ding{55} & $\checkmark$ \\
		\emph{LABELING-AND-DOCUMENTATION-03}\textsuperscript{*} & 2 & \ding{55} & 55,556 & 31,195 \\
		\emph{LABELING-AND-DOCUMENTATION-04 (!)}\textsuperscript{*} & \\
		\emph{LABELING-AND-DOCUMENTATION-05}\textsuperscript{*} & 39 & \ding{55} & \ding{55} & 978 \\
		\emph{LABELING-AND-DOCUMENTATION-06}\textsuperscript{*} &  302 & 46,718 & $\checkmark$ & $\checkmark$ \\
    \bottomrule
    \end{tabular}
    \caption{Thesauri Evaluation - Labeling and Documentation (2)}
		\label{tab:thesauri-evaluation-labeling-and-documentation-2}
    \end{center}
\end{table}

\begin{table}[H]
    \begin{center}
    \begin{tabular}{@{}lcccccccccccc@{}}
           & \multicolumn{12}{c}{\textbf{Data Sets}}
    \\  \cmidrule{2-13}
    \\       \textbf{Structure}
           & \emph{\rot{TheSoz}}
           & \emph{\rot{STW}}
           & \emph{\rot{AGROVOC}}
					 & \emph{\rot{TGN}}
           & \emph{\rot{UNESCO}}
					 & \emph{\rot{ODT}}
					 & \emph{\rot{SSWT}}
					 & \emph{\rot{GBA-GU}}
					 & \emph{\rot{GBA-GTS}}
					 & \emph{\rot{GBA-L}}
					 & \emph{\rot{GBA-LU}}
					 & \emph{\rot{CECCT}}
    \\ \midrule
		\emph{STRUCTURE-01}\textsuperscript{**} & 1 & 1,074 & $\checkmark$ & $\checkmark$ & 1 & 5 & 1 & $\checkmark$ & $\checkmark$ & $\checkmark$ & $\checkmark$ & $\checkmark$ \\
		\emph{STRUCTURE-02 (!)}\textsuperscript{*} &  &  &  &  &  & \\
		\emph{STRUCTURE-03}\textsuperscript{**} & $\checkmark$ & $\checkmark$ & $\checkmark$ & $\checkmark$ & 84 & $\checkmark$ & $\checkmark$ & $\checkmark$ & $\checkmark$ & $\checkmark$ & $\checkmark$ & $\checkmark$ \\
		\emph{STRUCTURE-04}\textsuperscript{*} & 2,906 & 8,046 & 726 & $\checkmark$ & 3,840 & 12 & 124 & 84 & 256 & 68 & 22 & 2,422 \\
		\emph{STRUCTURE-05}\textsuperscript{*} & $\checkmark$ & $\checkmark$ & $\checkmark$ & $\checkmark$ & \ding{55} & 90 & 5,150 & $\checkmark$ & $\checkmark$ & $\checkmark$ & $\checkmark$ & 9,864 \\
		\emph{STRUCTURE-06}\textsuperscript{*} & 1,457 & 37 & $\checkmark$ & $\checkmark$ & \ding{55} & $\checkmark$ & 4 & 1 & 1 & 64 & $\checkmark$ & 136 \\
		\emph{STRUCTURE-07}\textsuperscript{**} & 40 & 5,370 & $\checkmark$ & $\checkmark$ & \ding{55} & $\checkmark$ & $\checkmark$ & $\checkmark$ & $\checkmark$ & $\checkmark$ & $\checkmark$ & $\checkmark$ \\
		\emph{STRUCTURE-08 (!)}\textsuperscript{***} &  &  &  &  &  & \\
		\emph{STRUCTURE-09}\textsuperscript{*} & 7,897 & 19,844 & 99 & $\checkmark$ & 552 & 2 & 16 & 26 & $\checkmark$ & $\checkmark$ & $\checkmark$ & 82 \\
		\emph{STRUCTURE-10}\textsuperscript{**} & $\checkmark$ & $\checkmark$ & $\checkmark$ & $\checkmark$ & $\checkmark$ & $\checkmark$ & $\checkmark$ & $\checkmark$ & $\checkmark$ & $\checkmark$ & $\checkmark$ & $\checkmark$ \\
    \bottomrule
    \end{tabular}
    \caption{Thesauri Evaluation - Structure (1)}
		\label{tab:thesauri-evaluation-structure-1}
    \end{center}
\end{table}

\begin{table}[H]
    \begin{center}
    \begin{tabular}{@{}lcccc@{}}
           & \multicolumn{4}{c}{\textbf{Data Sets}}
    \\  \cmidrule{2-5}
    \\       \textbf{Structure}
					 & \emph{\rot{EARTh}}
					 & \emph{\rot{GEMET}}
					 & \emph{\rot{EuroVoc}}
					 & \emph{\rot{SLD}}
    \\ \midrule
		\emph{STRUCTURE-01}\textsuperscript{**} & 18,240 & \ding{55} & 55,757 & 31,195 \\
		\emph{STRUCTURE-02 (!)}\textsuperscript{*} & \\
		\emph{STRUCTURE-03}\textsuperscript{**} & 39 & 4,244 & $\checkmark$ & $\checkmark$ \\
		\emph{STRUCTURE-04}\textsuperscript{*} & 11,286 & 74 & $\checkmark$ & $\checkmark$ \\
		\emph{STRUCTURE-05}\textsuperscript{*} & $\checkmark$ & \ding{55} & $\checkmark$ & $\checkmark$ \\
		\emph{STRUCTURE-06}\textsuperscript{*} & 239,346 & \ding{55} & 13,876 & $\checkmark$ \\
		\emph{STRUCTURE-07}\textsuperscript{**} & 110,015 & \ding{55} & 366,155 & 155,975 \\
		\emph{STRUCTURE-08 (!)}\textsuperscript{***} & \\
		\emph{STRUCTURE-09}\textsuperscript{*} & 107,195 & 32 & $\checkmark$ & $\checkmark$ \\
		\emph{STRUCTURE-10}\textsuperscript{**} & 27 & 2,122 & $\checkmark$ & $\checkmark$ \\
    \bottomrule
    \end{tabular}
    \caption{Thesauri Evaluation - Structure (2)}
		\label{tab:thesauri-evaluation-structure-2}
    \end{center}
\end{table}

\begin{table}[H]
    \begin{center}
    \begin{tabular}{@{}lcccccccccccc@{}}
           & \multicolumn{12}{c}{\textbf{Data Sets}}
    \\  \cmidrule{2-13}
    \\       \textbf{Language Tag Cardinality}
           & \emph{\rot{TheSoz}}
           & \emph{\rot{STW}}
           & \emph{\rot{AGROVOC}}
					 & \emph{\rot{TGN}}
           & \emph{\rot{UNESCO}}
					 & \emph{\rot{ODT}}
					 & \emph{\rot{SSWT}}
					 & \emph{\rot{GBA-GU}}
					 & \emph{\rot{GBA-GTS}}
					 & \emph{\rot{GBA-L}}
					 & \emph{\rot{GBA-LU}}
					 & \emph{\rot{CECCT}}
    \\ \midrule
		\emph{LANGUAGE-TAG-CARDINALITY-01}\textsuperscript{**} & 9,435 & 13,468 & 98,894 & $\checkmark$ &  & 541 & 10,147 & 5,117 & 2,061 & 1,742 & 2,272 & 15,550 \\
		\emph{LANGUAGE-TAG-CARDINALITY-02}\textsuperscript{*} & 8,222 & 36,936 & \ding{55} & $\checkmark$ &  & 265 & 3,627 & 2,212 & 635 & 631 & 1,253 & 9,607 \\
		\emph{LANGUAGE-TAG-CARDINALITY-03}\textsuperscript{*} & 8,222 & $\checkmark$ & 135 & $\checkmark$ &  & $\checkmark$ & $\checkmark$ & $\checkmark$ & $\checkmark$ & $\checkmark$ & $\checkmark$ & $\checkmark$ \\
		\emph{LANGUAGE-TAG-CARDINALITY-04}\textsuperscript{*} & $\checkmark$ & 476 & \ding{55} & 50 &  & $\checkmark$ & $\checkmark$ & $\checkmark$ & $\checkmark$ & $\checkmark$ & $\checkmark$ & $\checkmark$ \\
    \bottomrule
    \end{tabular}
    \caption{Thesauri Evaluation - Language Tag Cardinality (1)}
		\label{tab:thesauri-evaluation-language-tag-cardinality-1}
    \end{center}
\end{table}

\begin{table}[H]
    \begin{center}
    \begin{tabular}{@{}lcccc@{}}
           & \multicolumn{4}{c}{\textbf{Data Sets}}
    \\  \cmidrule{2-5}
    \\       \textbf{Language Tag Cardinality}
					 & \emph{\rot{EARTh}}
					 & \emph{\rot{GEMET}}
					 & \emph{\rot{EuroVoc}}
					 & \emph{\rot{SLD}}
    \\ \midrule
		\emph{LANGUAGE-TAG-CARDINALITY-01}\textsuperscript{**} & \ding{55} & 2,318,895 & \ding{55} & 30,781 \\
		\emph{LANGUAGE-TAG-CARDINALITY-02}\textsuperscript{*} & \ding{55} & \ding{55} & \ding{55} & \ding{55} \\
		\emph{LANGUAGE-TAG-CARDINALITY-03}\textsuperscript{*} & 224,206 & \ding{55} & \ding{55} & 31,195 \\
		\emph{LANGUAGE-TAG-CARDINALITY-04}\textsuperscript{*} & \ding{55} & \ding{55} & $\checkmark$ & $\checkmark$ \\
    \bottomrule
    \end{tabular}
    \caption{Thesauri Evaluation - Language Tag Cardinality (2)}
		\label{tab:thesauri-evaluation-language-tag-cardinality-2}
    \end{center}
\end{table}

\begin{table}[H]
    \begin{center}
    \begin{tabular}{@{}lcccccccccccc@{}}
           & \multicolumn{12}{c}{\textbf{Data Sets}}
    \\  \cmidrule{2-13}
    \\       \textbf{Constraints}
           & \emph{\rot{TheSoz}}
           & \emph{\rot{STW}}
           & \emph{\rot{AGROVOC}}
					 & \emph{\rot{TGN}}
           & \emph{\rot{UNESCO}}
					 & \emph{\rot{ODT}}
					 & \emph{\rot{SSWT}}
					 & \emph{\rot{GBA-GU}}
					 & \emph{\rot{GBA-GTS}}
					 & \emph{\rot{GBA-L}}
					 & \emph{\rot{GBA-LU}}
					 & \emph{\rot{CECCT}}
    \\ \midrule
		\emph{PROPERTY-DOMAIN-01 (!)}\textsuperscript{***} &  &  &  &  &  & \\
		\emph{PROPERTY-RANGES-01 (!)}\textsuperscript{***} &  &  &  &  &  & \\
		\emph{DISJOINT-PROPERTIES-01 (!)}\textsuperscript{***} &  &  &  &  &  & \\
		\emph{DISJOINT-PROPERTIES-02 (!)}\textsuperscript{***} &  &  &  &  &  & \\
		\emph{DISJOINT-CLASSES-01 (!)}\textsuperscript{***} &  &  &  &  &  & \\
		\emph{EQUIVALENT-PROPERTIES-01 (!)}\textsuperscript{*} &  &  &  &  &  & \\
		\emph{UNIVERSAL-QUANTIFICATIONS-01 (!)}\textsuperscript{***} &  &  &  &  &  & \\
		\emph{CONTEXT-SPECIFIC-VALID-CLASSES-01 (!)}\textsuperscript{*} &  &  &  &  &  & \\
		\emph{CONTEXT-SPECIFIC-VALID-PROPERTIES-01 (!)}\textsuperscript{*} &  &  &  &  &  & \\
		\emph{RECOMMENDED-PROPERTIES-01 (!)}\textsuperscript{*} &  &  &  &  &  & \\
		\emph{VOCABULARY-01 (!)}\textsuperscript{***} &  &  &  &  &  & \\
		\emph{HTTP-URI-SCHEME-VIOLATION (!)}\textsuperscript{***} \\
    \bottomrule
    \end{tabular}
    \caption{Thesauri Evaluation - Constraints (1)}
		\label{tab:thesauri-evaluation-constraints-1}
    \end{center}
\end{table}

\begin{table}[H]
    \begin{center}
    \begin{tabular}{@{}lcccc@{}}
           & \multicolumn{4}{c}{\textbf{Data Sets}}
    \\  \cmidrule{2-5}
    \\       \textbf{Constraints}
					 & \emph{\rot{EARTh}}
					 & \emph{\rot{GEMET}}
					 & \emph{\rot{EuroVoc}}
					 & \emph{\rot{SLD}}
    \\ \midrule
		\emph{PROPERTY-DOMAIN-01 (!)}\textsuperscript{***}\\
		\emph{PROPERTY-RANGES-01 (!)}\textsuperscript{***}\\
		\emph{DISJOINT-PROPERTIES-01 (!)}\textsuperscript{***}\\
		\emph{DISJOINT-PROPERTIES-02 (!)}\textsuperscript{***}\\
		\emph{DISJOINT-CLASSES-01 (!)}\textsuperscript{***}\\
		\emph{EQUIVALENT-PROPERTIES-01 (!)}\textsuperscript{*}\\
		\emph{UNIVERSAL-QUANTIFICATIONS-01 (!)}\textsuperscript{***}\\
		\emph{CONTEXT-SPECIFIC-VALID-CLASSES-01 (!)}\textsuperscript{*} \\
		\emph{CONTEXT-SPECIFIC-VALID-PROPERTIES-01 (!)}\textsuperscript{*} \\
		\emph{RECOMMENDED-PROPERTIES-01 (!)}\textsuperscript{*} \\
		\emph{VOCABULARY-01 (!)}\textsuperscript{***} \\
		\emph{HTTP-URI-SCHEME-VIOLATION (!)}\textsuperscript{***} \\
    \bottomrule
    \end{tabular}
    \caption{Thesauri Evaluation - Constraints (2)}
		\label{tab:thesauri-evaluation-constraints-2}
    \end{center}
\end{table}

\section{Evaluation of Metadata on Statistical Classifications (XKOS)}

As part of future work, the quality of metadata on statistical classifications (XKOS) data sets will be evaluated by validating appropriate RDF constraints assigned to several RDF constraint types.

\subsection{Data Sets Overview}

\begin{table}[H]
	\centering
		\begin{tabular}{l|l}
      \textbf{Abbr.} & \textbf{Statistical Classifications} \\		
      \hline
    \emph{NAF} & \emph{Nomenclature d'activités française}\tablefootnote{\url{http://rdf.insee.fr/codes/index.html}} \\
		\emph{PCS} & \emph{Nomenclature des Professions et Catégories Socioprofessionnelles}\tablefootnote{\url{http://rdf.insee.fr/codes/index.html}} \\
		\emph{CJ} & \emph{Nomenclature des catégories juridiques}\tablefootnote{\url{http://rdf.insee.fr/codes/index.html}} \\
		\emph{ISIC} & \\
		\emph{ISCO} & \emph{International Standard Classification of Occupations} \\
		\end{tabular}
	\caption{Statistical Classifications Abbreviations}
	\label{tab:statistical-classifications-abbreviations}
\end{table}

\emph{Nomenclature d'activités française (NAF)} is the French refinement of the \emph{NACE} classification expressed in XKOS having explanatory notes.
\emph{Nomenclature des Professions et Catégories Socioprofessionnelles (PCS)} and \emph{Nomenclature des catégories juridiques (CJ)} are French classifications expressed in XKOS.
The statistical classification \emph{ISIC} has explanatory notes too. 


\section{Conclusion}

We identified and published by today 81 types of constraints that are required by various stakeholders for data applications.
In close collaboration with several domain experts for the social, behavioral, and economic sciences (SBE), we formulated and implemented 115 constraints on three different vocabularies (DDI-RDF, QB, and SKOS) and classified them according to their severity level and  whether their type is expressible by different types of constraint languages - RDFS/OWL, high-level constraint languages, and SPARQL. 
Using these constraints, we evaluated  the data quality of 15,694 data sets (4.26 billion triples) of research data for the SBE sciences obtained from 33 SPARQL endpoints.

\bibliography{../../../literature/literature}{}
\bibliographystyle{plain}
\setcounter{tocdepth}{1}
\end{document}